\documentclass[manuscript,acmsmall,screen,nonacm]{acmart}

\AtBeginDocument{%
  }

\newcommand\blfootnote[1]{%
  \begingroup
  \renewcommand\thefootnote{}%
  \footnote{#1}%
  \addtocounter{footnote}{-1}%
  \endgroup
}

\usepackage{xurl}
\usepackage{amsmath,amssymb,amsfonts}
\usepackage{algorithm, algpseudocode}
\usepackage{subcaption}
\usepackage{booktabs}

\usepackage{graphicx}
\usepackage{textcomp}
\usepackage{pifont}
\usepackage{multirow}
\usepackage{flushend}
\usepackage{microtype}
\usepackage{rotating}
\usepackage{tikz}
\usepackage{listings}
\usepackage{makecell}
\usepackage{enumitem}
\usepackage{wasysym}
\usepackage[table]{xcolor}
\usepackage{tcolorbox}

\definecolor{heatHigh}{RGB}{180,230,180}   
\definecolor{heatNear}{RGB}{220,240,180}   
\definecolor{heatLow}{RGB}{255,220,180}    
\definecolor{heatZero}{RGB}{240,180,180}   
\newcommand{\HH}[2]{%
    \ifnum#1=100 \cellcolor{heatHigh}%
    \else\ifnum#1=98  \cellcolor{heatNear}%
    \else\ifnum#1=8   \cellcolor{heatLow}%
    \else             \cellcolor{heatZero}%
    \fi\fi\fi
    #2\%}

\definecolor{heatHigh}{RGB}{180,230,180}   
\definecolor{heatHi99}{RGB}{195,232,175}   
\definecolor{heatHi98}{RGB}{210,234,170}   
\definecolor{heatHi97}{RGB}{225,236,165}   
\definecolor{heatHi96}{RGB}{235,235,160}   
\definecolor{heatZero}{RGB}{240,180,180}   

\newcommand{\HHH}[2]{%
    \ifnum#1=100 \cellcolor{heatHigh}%
    \else\ifnum#1=99  \cellcolor{heatHi99}%
    \else\ifnum#1=98  \cellcolor{heatHi98}%
    \else\ifnum#1=97  \cellcolor{heatHi97}%
    \else\ifnum#1=96  \cellcolor{heatHi96}%
    \else             \cellcolor{heatZero}%
    \fi\fi\fi\fi\fi
    #2\%}

\let\oldtexttt\texttt
\renewcommand{\texttt}[1]{{\small\oldtexttt{#1}}}

\definecolor{funcblue}{RGB}{10,30,60}
\definecolor{argbrown}{RGB}{85,70,0}
\definecolor{retgray}{RGB}{40,40,40}

\definecolor{rowBaseline}{RGB}{240,240,240}
\definecolor{rowDefended}{RGB}{210,235,210}
\definecolor{blockedBg}{RGB}{220,80,80}
\definecolor{blockedFg}{RGB}{255,255,255}
\definecolor{partialBg}{RGB}{240,180,100}
\definecolor{summaryBg}{RGB}{225,225,245}

\newcommand{\BLOCKED}{%
    \cellcolor{blockedBg}\textcolor{blockedFg}{{\textsc{blocked}}}}
\newcommand{\PARTIAL}[1]{%
    \cellcolor{partialBg}$#1 -$ {\textsc{blocked}}}

\begin{document}

\title{Safeguarding LLMs Against Misuse and AI-Driven Malware Using Steganographic Canaries}

\author{Md Raz}
\email{md.raz@nyu.edu}

\author{Venkata Sai Charan Putrevu}
\email{v.putrevu@nyu.edu}

\author{Meet Udeshi}
\email{m.udeshi@nyu.edu}

\author{Prasanth Krishnamurthy}
\email{prashanth.krishnamurthy@nyu.edu}

\author{Farshad Khorrami}
\email{khorrami@nyu.edu}

\author{Ramesh Karri}
\email{rkarri@nyu.edu}
\affiliation{%
  \institution{Department of Electrical and Computer Engineering, Tandon School of Engineering, New York University}
  \city{Brooklyn}
  \state{New York}
  \country{USA}
  }

\blfootnote{This work was supported in part by the DOE NETL grants DE-CR0000051 and DE-CR0000017, the NSF SaTC grant 2039615, and NYSTAR C220160.}

\renewcommand{\shortauthors}{Raz et al.}

\begin{abstract}

    AI-powered malware increasingly exploits cloud-hosted generative-AI services and large language models (LLMs) as analysis engines for reconnaissance, file triage, and code generation. Simultaneously, routine enterprise uploads expose sensitive documents to third-party AI vendors. Both threats converge at the AI service ingestion boundary, yet existing defenses focus on endpoints and network perimeters, leaving organizations with limited visibility once plaintext reaches an LLM service.
    To address this, we present a framework based on steganographic canary files: realistic documents carrying cryptographically derived identifiers embedded via complementary encoding channels. A pre-ingestion filter extracts and verifies these identifiers before LLM processing, enabling passive, format-agnostic detection and data provenance without semantic classification. 
    We support two modes of operation where Mode~A marks existing sensitive documents with layered symbolic encodings (whitespace substitution, zero-width character insertion, homoglyph substitution), while Mode~B generates synthetic canary documents using linguistic steganography (arithmetic coding over GPT-2), augmented with compatible symbolic layers.
    We model increasing document pre-processing and adversarial capability for both modes via a four-tier transport-transform taxonomy:
    All methods achieve 100\% identifier recovery under benign and sanitization workflows (Tiers~1--2). The hybrid Mode~B maintains 97\% through targeted adversarial transforms (Tier~3) while symbolic-only Mode~A provides full coverage through Tier~2. We show that improper layer composition can reduce Tier~3 recovery from 97\% to 0\% via cross-layer interference, yielding empirical composition principles. 
    Verification via cryptographic identifiers (HMAC and EdDSA) produce zero false positives while surfacing deployment trade-offs. An end-to-end case study against an LLM-orchestrated ransomware pipeline confirms that both modes detect and block canary-bearing uploads during reconnaissance, before file encryption begins.
    To our knowledge, this is the first framework to systematically combine symbolic and linguistic text steganography into layered canary documents for detecting unauthorized LLM processing, evaluated against a transport-threat taxonomy tailored to AI malware.

\end{abstract}



\maketitle

\section{Introduction} \label{sec:introduction}
    AI-powered malware depend on cloud-hosted large language models (LLMs) as analysis engines. Autonomous agentic ransomware can now execute the full attack lifecycle, including reconnaissance, file targeting, data exfiltration, encryption, and ransom-note generation, by leveraging an external LLM without human control~\cite{newman2025airansomware}.
    This LLM-analysis dependency extends to AI-assisted espionage agents that triage exfiltrated documents for intelligence value~\cite{gtig2025adversarial}, automated credential and secret harvesters that use contextual understanding to extract API keys and tokens from configuration files, and social-engineering tools that analyze internal communications to craft targeted spear-phishing campaigns~\cite{microsoft2024stayingahead}. 
    In parallel, enterprise adoption of LLM tools has exposed a new data leakage channel where employees submit sensitive documents to third-party AI services, with reports of senior officials uploading marked government documents to public ChatGPT instances~\cite{swain2026cisa}.

    \begin{figure}[t]
        \centering
        \includegraphics[width=0.9\linewidth, trim=0 0.5mm 0 6mm, clip]{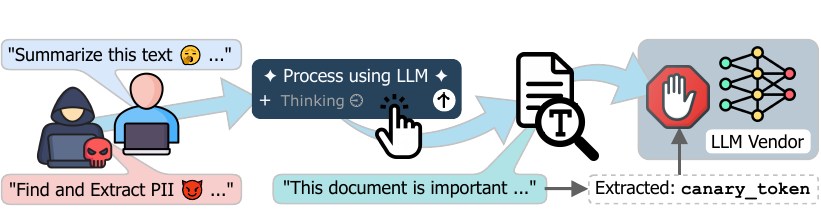}
        \caption{Overview: steganographic canary files detect unauthorized document submission to AI services at the ingestion boundary, before LLM processing occurs.}
        \label{fig:stego_intro}
    \end{figure}

    Current defenses offer limited protection once plaintext leaves an organization's boundary. This gap is acute for AI-powered malware: defenses are signature or behavior-based but LLM-assisted malware maintain a near-imperceptible footprint while conducting reconnaissance, file analysis, and payload generation via external models~\cite{raz2025ransomware30}. Once an adversary has obtained files, whether through compromise or authorized access, the question becomes whether the files can effectively be weaponized via an LLM. This creates a defensive opportunity at the processing boundary that conventional endpoint and perimeter protections do not cover, namely detecting unauthorized LLM analysis of documents that have already left the organization's control.
    
    Network Data Loss Prevention (DLP) and endpoint controls can block destinations or match patterns, but they are easily bypassed when data is uploaded through compromised accounts, personal devices, or encrypted channels. Access controls limit who can read a file, but they do not prevent an authorized (or compromised) user from leaking data. Semantic classifiers for PII and secrets add another layer consistent with emerging AI governance frameworks, but remain brittle in practice as domain-specific language can evade detection, while broad rules create costly false positives. Even when an organization suspects exposure, it is often impossible to determine whether a specific document was uploaded to a specific AI service as existing controls provide little file-level accountability after data crosses the boundary. Accordingly, OWASP's Top-10 for LLM Applications lists Sensitive Information Disclosure (LLM02:2025) as a top-tier risk~\cite{owasp2025llmtop10}. Our work addresses this core gap where there exists a lack of a portable, file-content-level tripwire that remains intact under common transformations and can be detected by the receiving AI service's ingestion boundary before model processing, without relying on semantic inspection of the contents.

    We propose a defensive framework for detecting unauthorized submission of controlled documents to LLM services. The framework uses plaintext canary files seeded across organizational file shares, where each canary carries a visually indistinguishable, cryptographically derived identifier steganographically embedded in the textual content itself. If a user or attacker uploads or pastes the document into a participating AI service, a vendor-side pre-ingestion filter extracts and verifies the identifier, then blocks or flags the submission before the LLM processes it. The defense operates as a second line that activates when traditional perimeter defenses have failed, as shown in Figure~\ref{fig:stego_intro}.
    We design framework with three practical properties. First, it provides content-level persistence as the embedded identifier resides in plaintext and travels with the content, surviving common handling such as copy/paste, reflow, and format conversion, while also supporting ingestion into LLM input pipelines. Second, it enables passive detection at the processing boundary (the AI vendor's ingestion boundary) rather than relying on callback-based honeytokens that can be blocked. Third, it implements orthogonal embedding classes that preserve visible content (symbolic encodings) or can survive character-level normalization (linguistic encodings). This layered design improves robustness to heterogeneous AI pre-processing pipelines using a ``defense-in-depth'' approach, where deeper layers provide higher chances of encoding survival.

    To evaluate the proposed framework, we structure our study around five research questions to examine feasibility, robustness, and deployment realism end to end:
    
    \begin{enumerate}[noitemsep,nolistsep,leftmargin=24pt]
        
        \item[\textbf{\textsf{RQ1}}] \textbf{Feasibility:} Can steganographic identifiers be embedded into plaintext canary files such that a per-file cryptographic identifier can be recovered after a representative transport and transformation pipeline?
        
        \item[\textbf{\textsf{RQ2}}] \textbf{Robustness:} How do symbolic and linguistic embeddings compare in resilience to benign handling, sanitization, and adversarial transformations, and what failure modes are exposed?
        
        \item[\textbf{\textsf{RQ3}}] \textbf{Defense-in-Depth:} Does layering multiple methods (e.g., symbolic embeddings on disjoint character surfaces or hybrid embeddings) provide measurably broader coverage across transport conditions than any single method or method class alone?
        
        \item[\textbf{\textsf{RQ4}}] \textbf{Practical Deployment:} What are the computational costs, textual capacity limits, and operational trade-offs of deploying steganographic canary files at organizational scale?
        
        \item[\textbf{\textsf{RQ5}}] \textbf{End-to-End Detection:} Can the complete pipeline reliably detect and block unauthorized uploads in a realistic simulated AI-driven ransomware workflow?
    
    \end{enumerate}

    \noindent Through empirical evaluations guided by these questions, we determine whether steganographic canary files with cryptographically derived embeddings are technically viable, robust against changes, and deployable in practice. We make the following key contributions:

    \begin{enumerate}[noitemsep,nolistsep,leftmargin=17pt]

        \item[\textbf{\textsf{C1}}] \textbf{Defensive framework:} We design a framework that embeds cryptographically verifiable identifiers via layered symbolic and linguistic steganographic channels, supporting both shared-key (HMAC-SHA256) and public-key (EdDSA/Ed25519) verification. \textbf{Mode~A} marks existing sensitive documents with symbolic encodings at sub-millisecond cost; \textbf{Mode~B} generates synthetic canary documents via arithmetic coding over GPT-2, augmented with compatible symbolic layers. The framework is grounded in a threat model that formalizes the AI-powered malware interception point, where adversaries who have obtained files must submit them to an external LLM for analysis, alongside incidental insider uploads.

        \item[\textbf{\textsf{C2}}] \textbf{Transport-transform taxonomy and systematic evaluation:} We present a four-tier transport-transform taxonomy for the LLM-upload threat model, with systematic evaluation of individual and stacked method robustness across seven configurations and six composite transport chains. 
        The evaluation reveals per-file binary recovery behavior, orthogonal failure surfaces, and layering rules demonstrating reduced survival rates for improper method compositions.

        \item[\textbf{\textsf{C3}}] \textbf{End-to-end validation:} We demonstrate the full canary file lifecycle in an end-to-end case study against a \textsc{PromptLock}-style LLM-orchestrated ransomware pipeline, detecting and blocking the attack during its reconnaissance phase before any file encryption occurs.

    \end{enumerate}

    The remainder of the paper is organized as follows. 
    Section~\ref{sec:background} presents the background and limitations of existing defenses for detecting unauthorized document submission to LLMs.
    Section~\ref{sec:threat_model} formalizes our threat model and assumptions, including the AI-powered malware categories that share the LLM-analysis choke-point.
    Section~\ref{sec:architecture} presents the steganographic canary file framework and its encode--transport--decode pipeline.
    Section~\ref{sec:experimental_setup} describes the evaluation methodology, datasets, transport transforms, and measurement criteria. 
    Section~\ref{sec:evaluation} reports feasibility, robustness, layering, and end-to-end case study results. 
    Section~\ref{sec:discussion} examines deployment considerations, limitations, and broader implications.
    Finally, Section~\ref{sec:conclusion} concludes.
              
\section{Background and Related Work} \label{sec:background}
    We position our work at the intersection of three research threads: the emerging class of AI-powered malware that depends on external LLMs for document analysis, data-protection and deception-based defenses, and text steganography. We close by comparing our work with prior art.

    \subsection{AI-Powered Adversaries}
        Cloud-hosted LLM interfaces that enable benign document analysis have simultaneously created a new operational dependency for adversaries. Greshake et al.\ showed that LLM-integrated applications blur the boundary between data and instructions as LLMs can exfiltrate user data via generated content through embedded adversarial prompts~\cite{greshake2023indirect}. A joint Microsoft--OpenAI disclosure later documented five nation-state groups and identified their use of ChatGPT for reconnaissance, vulnerability research, scripting, and social-engineering content generation~\cite{microsoft2024stayingahead}. Google's Threat Intelligence Group (GTIG) subsequently reported an escalation from productivity-oriented LLM misuse to runtime integration, detailing malware families such as \textsc{PromptFlux} and \textsc{PromptSteal}, both of which generate commands and code during execution~\cite{gtig2025adversarial}.
        
        This trend is especially consequential for ransomware. The \textsc{Ransomware~3.0} prototype (publicly identified by ESET as \textsc{PromptLock}) is the first fully closed-loop, LLM-orchestrated ransomware where a lightweight Go binary embeds natural-language prompts rather than pre-written attack code, delegates reconnaissance, file-system enumeration, payload generation, data exfiltration, encryption, and personalized extortion-note composition to an LLM, and produces polymorphic Lua scripts that adapt to each victim environment at runtime~\cite{raz2025ransomware30}. The malware cannot execute an intelligent attack without first ingesting information about the environment and candidate files. 

        Furthermore, the same LLM-analysis dependency extends beyond ransomware to multiple AI-powered malware categories. AI-assisted espionage agents can rapidly triage exfiltrated document troves for intelligence value, automated credential harvesters leverage contextual LLM understanding to extract secrets from configuration files and source code, and social-engineering tools analyze internal communications to craft targeted spear-phishing campaigns. All share a common operational requirement: submitting stolen documents to an external LLM for analysis. This shared interception point at the AI service ingestion boundary motivates our defensive approach~\cite{gtig2025adversarial, gupta2023threatgpt}.

    \subsection{The LLM Data Exfiltration Channel}
        Beyond adversarial exploitation, enterprise adoption of LLMs has created a channel for sensitive data leakage. In March 2023, Samsung Electronics employees on multiple occasions pasted proprietary company materials such as semiconductor source code into ChatGPT, prompting a company-wide ban on all generative-AI services~\cite{gurman2023samsung}. Cyberhaven Labs' telemetry across 7~million knowledge workers shows that enterprise AI usage grew 4.6$\times$ year-over-year, with 34.8\% of data submitted to AI tools classified as sensitive, and 83.8\% of enterprise AI traffic flows to external tools rated medium, high, or critical risk~\cite{cyberhaven2025adoption}.
        As a result of LLM-assisted workflows normalizing the transfer of plaintext to third-party services for summarization, extraction, and code generation, benign productivity use and deliberate exfiltration now share the same interface and transport path.

    \subsection{Existing Defenses and Limitations}
        Existing defenses leave a clear gap at the AI service processing boundary, especially in the context of agentic malware and benign-looking workflows. 
        
        \begin{itemize}[noitemsep,nolistsep,leftmargin=*]
        
            \item \textbf{Endpoint controls are channel-centric:} Network endpoint monitors inspect traffic, destinations, or host behavior, and work best when exfiltration paths and signatures are known~\cite{alneyadi2016dlp}. In the LLM setting, however, data may leave through personal devices, browser copy/paste, or encrypted tunnels (i.e. in-content rather than channel), often resembling legitimate use~\cite{homoliak2019insiderthreat}.
            
            \item \textbf{Access control ineffective after compromise:} File permissions limit who can read a document, but once a user (or compromised session) has read access, existing controls generally do not prevent copying, reformatting, or uploading the content to an external AI service~\cite{homoliak2019insiderthreat}.
            
            \item \textbf{Semantic inspection is brittle/costly:} PII detectors and policy-based classifiers can help, but rule-based methods are easy to evade via targeted re-wording or jargon-based obfuscation, while semantic models produce higher false positives and often require inspection of file contents~\cite{mishra2025pii}.

            \item \textbf{AI-malware defenses miss the analysis channel:} Traditional detection relies on process, filesystem, and behavioral indicators~\cite{cen2024ransomware,anand2025larm}. AI-powered malware can reduce these signals via polymorphic code, task decomposition, and low-and-slow disk access patterns~\cite{gupta2023threatgpt,raz2025ransomware30} while outsourcing analysis of files to external LLMs through channels unprotected by endpoint defenses. 

            \item \textbf{LLM safety guardrails are insufficient:} Modern LLMs employ trained refusal mechanisms for malicious requests, but these guardrails are routinely bypassed via prompt injection~\cite{greshake2023indirect}, jailbreaks, and multi-prompt decomposition where AI malware decomposes attacks across discrete prompts such that no single prompt appears overtly malicious~\cite{raz2025ransomware30}. Guardrails are provider-specific and version-dependent, making them unreliable as security controls. Our defense operates on document content at the ingestion boundary before content reaches the model, and functions regardless of model refusal behavior.            
            \item \textbf{Post-incident Forensics and attribution:} Even when exposure is suspected, organizations often cannot determine provenance of a document uploaded to a specific AI service, or whether it was transformed before submission, both of which aid incident response and investigation~\cite{casino2022forensics}. 

        \end{itemize}

    \subsection{Canary Files and Deception-Based Defenses}

        Deception-based defenses detect compromise by planting monitored artifacts whose unauthorized access signals attacker activity. Spitzner introduced this paradigm through honeypots and honeytokens~\cite{spitzner2003honeypots}, with later work extending it to filesystem honeyfiles~\cite{yuill2004honeyfiles} and beaconed documents that trigger hidden HTTP/DNS callbacks when opened outside the enterprise perimeter~\cite{bowen2009baiting}. Surveys cover these mechanisms across network, host, and data layers~\cite{han2018deception,zhang2021deception,putrevu2024comprehensive}.
        Modern deployments such as \textsc{Thinkst Canary}/\textsc{Canarytokens} operationalize this model using tripwires that alert via network callbacks~\cite{thinkst2015canary}. This works well when the exfiltration path preserves content, but fails when transport is offline, anonymized, or mediated by services that sanitize macros or active payloads, which is typical for LLM ingestion. 
 
     \subsection{Text Steganography}
        Text steganography provides the technical mechanism underlying our framework design, which embeds hidden information into textual content while preserving visible utility. Our framework draws from character-level symbolic steganography and generative linguistic steganography, and then combines them for the defense-in-depth paradigm. 

        \subsubsection{Symbolic steganography}

            \textbf{(i)~Whitespace substitution (WS)} replaces ordinary spaces and line breaks with visually equivalent Unicode whitespace characters to encode arbitrary bytes. \textsc{Innamark} formalized this as a Kotlin multiplatform library that embeds payloads in inter-word gaps~\cite{hellmeier2025innamark}, and the earlier \textsc{AITSteg} employed similar substitutions for covert messaging on social-media platforms~\cite{ahvanooey2018aitsteg}. 
            \textbf{(ii)~Zero-width character insertion (ZW)} uses non-printing code points between visible glyphs. \textsc{StegCloak} compresses and HMAC-encrypts a secret before encoding it as a sequence of six zero-width characters, achieving practical invisibility in browsers and messaging applications~\cite{kurolabs2020stegcloak}. 
            \textbf{(iii)~Homoglyph substitution (HG)} replaces Latin glyphs with visually confusable code points from other scripts such as Cyrillic or Greek, using mappings cataloged in the Unicode Technical Standard\,\#39~\cite{unicode2025uts39}. Boucher et al.\ showed that such imperceptible replacements can evade NLP classifiers, spam filters, and toxicity detectors~\cite{boucher2022badcharacters}. 
            These sub-types exhibit complementary failure modes where whitespace and zero-width methods offer high capacity but are fragile to normalization, and  homoglyph substitution is more robust to whitespace stripping but may be flagged by confusable-character scanners or Unicode security-aware steganalysis. This complementarity motivates multilayer symbolic encoding for more robust defense.
        
        \subsubsection{Linguistic steganography}
            The \textbf{Linguistic Method (LM)}, encodes hidden bits by constraining language-model-generated text, so that the output remains coherent while carrying a recoverable payload. 
            Neural linguistic steganography became practical when Ziegler et al.\ coupled arithmetic coding with GPT-2, achieving several bits encoded per token~\cite{ziegler2019neural}. Dai and Cai concurrently proposed patient-Huffman coding with formal near-imperceptibility guarantees~\cite{dai2019towards}. Subsequent work improved both throughput and security, as Self-Adjusting Arithmetic Coding (\textsc{SAAC})~dynamically tunes the truncation parameter at each decoding step, improving embedding rate by 15.3\% and KL divergence by 38.9\% over fixed baselines~\cite{shen2020saac}, \textsc{ADG} recursively partitions the vocabulary for provably secure encoding~\cite{zhang2021adg}, \textsc{Meteor} provides a symmetric-key stateful protocol for variable-entropy channels~\cite{kaptchuk2021meteor}, and \textsc{Discop} samples from ``distribution copies'' to maximize throughput while maintaining the exact cover distribution~\cite{ding2023discop}. A practical constraint shared by all linguistic methods is decoder synchronization, where the encoder and decoder must use an identical LLM. In our implementation we select GPT-2 (124\,M) as the model, accepting lower capacity and vocabulary in exchange for portability and reproducibility.

        \subsubsection{File-Format-level and Output Marking}
            Document formats such as PDF and OOXML provide embedding channels in container structures such as metadata, internal XML, revision fields, and embedded objects~\cite{castiglione2011ooxml}. Such format-coupled marks are typically lost under plaintext extraction and copy/paste, which dominate many LLM workflows. In contrast, LLM output watermarking embeds provenance into model-generated text (e.g., token-bias watermarking~\cite{kirchenbauer2023watermark} and production-scale \textsc{SynthID-Text}~\cite{dathathri2024synthid}) to establish whether a given text was AI-generated. We address the inverse problem of input-side content provenance where we mark pre-existing (or newly-generated) documents to detect unauthorized input ingestion (rather than output) by an AI service.

        \begin{table}[t]
            \centering
            \small
            \caption{Comparison of steganographic canaries (This work) with existing defenses:}
            \label{tab:comparison}
            \resizebox{0.78\textwidth}{!}{
            \setlength\extrarowheight{-4pt}
            \begin{tabular}{@{}lcccc@{}}
            \toprule
            \textbf{Property} & \textbf{Endpoint Protection} & \textbf{Active tokens} & \textbf{Doc.\ watermark} & \textbf{This Work} \\
            \midrule
            Channel-agnostic       & \ding{55} & \ding{55} & \ding{51} & \ding{51} \\
            Survives copy-paste    & \ding{55} & \ding{55} & \ding{55} & \ding{51} \\
            Passive (no callback)  & ---       & \ding{55} & \ding{51} & \ding{51} \\
            Format-agnostic        & \ding{55} & \ding{55} & \ding{55} & \ding{51} \\
            Works offline / Tor    & \ding{55} & \ding{55} & \ding{51} & \ding{51} \\
            \bottomrule
            \end{tabular}}
        \end{table}

    \subsection{Positioning Our Work}

        Steganographic canaries address a gap not covered by existing defenses, namely a passive, portable, file-content-level tripwire that survives common transformations and can be detected at the AI processing boundary without semantic inspection or classification. We summarize this distinction against other methods in Table~\ref{tab:comparison}.
        Compared with traditional endpoint defenses and DLP, steganographic canaries are channel-agnostic as the identifier travels with the text itself rather than a monitored network path. They also operate as a second line that activates at the processing boundary when network perimeter defenses have already been bypassed. Compared with active canary tokens, they require no callback and remain effective when active content is stripped or outbound signaling is blocked. Compared with document-level watermarking, they survive plaintext extraction and copy/paste into LLM interfaces. Compared with LLM output watermarking, they solve the inverse problem of detecting unauthorized input ingestion of protected content. 
        To our knowledge, no prior work has combined layered text-steganographic encodings (symbolic and linguistic) into canary documents and evaluated their survivability under a tiered transport-threat model tailored to the AI-powered malware threats. We formulate and evaluate this paradigm in this paper.                
\section{Threat Model} \label{sec:threat_model}

\begin{figure}[t]
        \centering
        \includegraphics[width=1\linewidth]{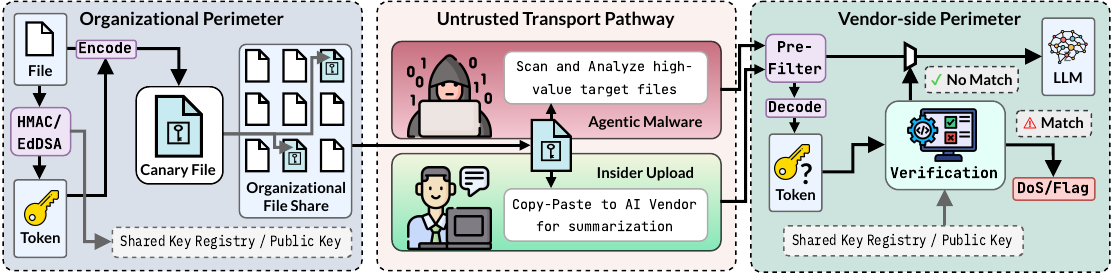}
        \caption{Threat model scenario showing the two motivating threat pathways (AI adversaries and insider/incidental upload), along with the framework encompassing the seeded canary files within the organizational boundary and steganographic identifier extraction / verification within the vendor-side detection boundary.}
        \label{fig:threat-model}
    \end{figure}

    We formalize the scenario (Figure~\ref{fig:threat-model}) in which steganographic canary files operate, the adversary capabilities they must withstand, and the transport transforms for robustness evaluation. Our threat model is motivated by two increasingly common AI security realities: (i)~AI-powered malware, including agentic ransomware, AI-assisted espionage agents, automated credential harvesters, and social-engineering tools, depend on external LLMs to analyze stolen files, creating a processing boundary detection opportunity that endpoint defenses do not cover; and (ii)~routine enterprise LLM adoption normalizes the sensitive plaintext submission to third-parties, blurring the boundary between legitimate use and data exfiltration~\cite{cyberhaven2025adoption}.
    A key assumption is that the adversary has already obtained the files, whether through compromise or authorized access. The proposed defense does not attempt to prevent exfiltration at the network perimeter (already addressed by traditional DLP), but rather focuses on the processing boundary point at which plaintext enters an external AI service for analysis. At this interception point, we ask whether content-level invisible identifiers can provide reliable detection, activating when perimeter defenses have failed.

    \subsection{Scenario} \label{sec:scenario}

        An organization (enterprise, government agency, research lab) maintains file shares containing sensitive plaintext artifacts such as source code, internal documentation, and configuration files. The organization seeds these shares with steganographic canary files, which are realistic documents indistinguishable from legitimate content under casual inspection, carrying a cryptographically derived identifier embedded via one or more channels. We detail the embedding methods, identifier construction, and deployment modes in Section~\ref{sec:architecture}, and identify two motivating pathways:

        \begin{itemize}[noitemsep,nolistsep,leftmargin=*]
        
            \item \textbf{AI-Powered Adversaries:} Agentic malware or an external attacker with AI-assistance gains user-space access to file shares through compromise, lateral movement, or supply-chain abuse. The adversary submits stolen files to an external LLM for analysis, credential discovery, summarization, or code/payload generation. Multiple AI-powered malware categories share this LLM-analysis dependency: (a)~LLM-orchestrated ransomware that analyzes files for extortion leverage~\cite{raz2025ransomware30}, (b)~AI-assisted espionage agents that rapidly analyze exfiltrated documents for intelligence value~\cite{gtig2025adversarial}, (c)~automated credential and secret harvesters that use contextual LLM understanding to extract API keys and tokens from configuration files, and (d)~social-engineering tools that analyze internal enterprise communications to craft targeted campaigns~\cite{microsoft2024stayingahead}. Because all four categories depend on ingesting stolen documents via an external LLM, the AI service ingestion boundary becomes a shared interception point and detection opportunity that is absent from traditional endpoint defenses.
            
            \item \textbf{Incidental or Insider Upload:} An authorized user, acting negligently or with malicious intent, copies or uploads sensitive content to a third-party AI service for summarization, code review, or other analysis. This pathway shares the same transport interface as adversarial exfiltration, making policy enforcement and forensic attribution challenging via conventional controls.
            
        \end{itemize}
        
        \noindent In both cases, an AI service (or an enterprise-controlled monitoring proxy) runs an ingest-time extraction filter over inbound plaintext, checking extracted values against a registry/manifest of known tokens. If a valid identifier is recovered, a configurable response is triggered (e.g., lockdown or breach alert). We detail the extraction and verification mechanism in Section~\ref{sec:architecture}, while the notification protocol and key-distribution model are considered out of scope.

    \subsection{Adversary Model}
    
        We model an adversary whose objective is to access and analyze sensitive plaintext using external AI services. The adversary controls content transport and pre-processing before submission but does not possess the organization's secret key or embedding configuration. We assume that the adversary may know the general defense class and embedding method families, but not the specific method selection, parameters, or key material used in a given deployment.
        
        We define escalating adversary capabilities from basic file access and upload capability up to active countermeasures. At the base level, the adversary can read and copy files from organizational file shares via insider privileges, compromised accounts, malware, or lateral movement, and may select which files to exfiltrate. The adversary can then submit file contents to external AI services via browser upload, copy/paste, or programmatic API calls, targeting complete documents or partial excerpts that contain the hidden identifier. The upload path may also apply transformations that unintentionally corrupt embedded channels including Unicode normalization, whitespace collapsing, smart-quote substitution, line reflow, or format-character stripping (see Table~\ref{tab:taxonomy}). At the upper bound, a sophisticated adversary suspecting steganographic embedding can apply targeted sanitization, trading overhead and content fidelity for reduced detection probability.

    \subsection{Transport-Transform Taxonomy} \label{sec:taxonomy}

        We organize the transforms an insider/adversary may accidentally or intentionally apply into four tiers where tier subsumes all transforms in lower tiers. Table~\ref{tab:taxonomy} lists the twelve individual transforms and Table~\ref{tab:chains} defines the composite chains used experimentally. We find that one structural property, validated empirically in Section~\ref{sec:evaluation}, is that no single Tier~1--3 transform destroys all embedding methods simultaneously, causing Tier~4 semantic rewriting to be the only tier that defeats all methods. Furthermore, we identify that chained ordering can also produce interaction effects. For example, Unicode Normalization Form KC (NFKC) normalization in Tier-2 may inadvertently preserve signals that a later Tier-3 strip would otherwise destroy. 
        
        \begin{table}[t]
            \centering
            \small
            \caption{Transport-transform taxonomy. Tiers reflect increasing adversary sophistication and intent. Each tier subsumes all transforms in lower tiers.}
            \label{tab:taxonomy}
            \resizebox{0.8\textwidth}{!}{
            \setlength\extrarowheight{-4pt}
            \begin{tabular}{@{}clll@{}}
            \toprule
            \textbf{ID} & \textbf{Transform} & \textbf{Tier} & \textbf{Typical Source} \\
            \midrule
            T00 & None                        & 0 \;--\;Passthrough & Baseline \\
            \midrule
            T01 & Copy-paste normalization    & 1\;--\;Benign & Clipboard, browser \\
            T02 & Line reflow                 & 1 & Editor word-wrap \\
            T03 & Smart-quote replacement     & 1 & Word processor, CMS \\
            T04 & Trailing-whitespace strip   & 1 & Editor, version control hooks \\
            \midrule
            T05 & Unicode NFKC normalization  & 2\;--\;Sanitization & Input validation \\
            T06 & Whitespace collapse (ASCII) & 2 & HTML/API ingest \\
            T07 & Format-character strip (Cf) & 2 & Sanitizer, web application firewall \\
            \midrule
            T08 & Zero-width char stripping   & 3\;--\;Adversarial & Targeted sanitizer \\
            T09 & Homoglyph normalization     & 3 & Confusable folding \\
            T10 & Full ASCII strip (Transliteration)           & 3 & Remove all non-ASCII \\
            \midrule
            T11 & Punctuation/case strip      & 4\;--\;Special & Content normalization \\
            T12 & LLM-based paraphrase        & 4 & Intentional rewrite \\
            \bottomrule
            \end{tabular}}
        \end{table}

        \begin{description}[style=unboxed,leftmargin=0pt,nosep]
        
            \item[\textbf{Tier~1\;--\;Benign / Incidental:}] Standard transforms applied by platforms, editors, and clipboard managers: line-ending normalization, reflow, smart-quote replacement, and trailing-whitespace stripping. These represent the minimum adversity any embedding must tolerate.
            
            \item[\textbf{Tier~2\;--\;Standard Sanitization:}] Security-conscious input processing that is not watermark-aware but damages specific encoding surfaces: NFKC normalization (T05), whitespace collapsing (T06), and format-character stripping (T07).
            
            \item[\textbf{Tier~3\;--\;Targeted Adversarial:}] An adversary suspecting steganographic embedding applies directed stripping: zero-width character removal (T08), confusable normalization (T09), or full non-ASCII stripping (T10).
            
            \item[\textbf{Tier~4\;--\;Special Transforms:}] Punctuation/case stripping (T11) and LLM-based paraphrase (T12) exhibit qualitatively different failure patterns from Tiers~1--3, producing inverse coverage profiles affecting both tokenization and symbols across embedding classes.

        \end{description}

        \begin{table}[t]
            \centering
            \small
            \caption{Composite transport chains used as experimental variables. Each chain applies its constituent transforms in sequence. T11 is tested individually but excluded from chaining because T11$\to$T12 is redundant.}
            \label{tab:chains}
            \resizebox{0.67\textwidth}{!}{
            \setlength\extrarowheight{-4pt}
            \begin{tabular}{@{}lll@{}}
            \toprule
            \textbf{Chain} & \textbf{Composition} & \textbf{Models} \\
            \midrule
            Tier-0  & T00 (Passthrough)                 & Baseline / control \\
            Tier-1  & T01 $\to$ T02 $\to$ T03 $\to$ T04  & User copy-paste \\
            Tier-2  & T05 $\to$ T06 $\to$ T07            & Platform sanitization \\
            Tier-3  & T08 $\to$ T09 $\to$ T10            & Steganography-aware attacker \\
            Tier-1+2 & Tier-1 $\to$ Tier-2               & Combined incidental processing \\
            Tier-1+2+3 & Tier-1 $\to$ Tier-2 $\to$ Tier-3 & Maximum non-semantic pipeline \\
            Tier-4  & T12                                 & LLM semantic rewriting \\
            \bottomrule
            \end{tabular}}
        \end{table}

    \subsection{Security Goals}
    
        Our security goals align with the intended role of canary files as file-level tripwires with content-level embeddings detectable at an AI service boundary. Embedded identifiers must be recoverable with high probability under incidental processing (Tier~1) and common sanitization (Tier~2), with layered embeddings to provide graceful degradation at higher tiers. Detection succeeds if any surviving layer yields a verifiable token, while also appearing benign under casual inspection. We do not claim indistinguishability against targeted statistical tests, as the defense relies instead on volume and the cost asymmetry of per-file analysis. Finally, recovered identifiers must be integrity-checked with negligible false positives.

    \subsection{Trust Assumptions \& Scope}

        \begin{description}[style=unboxed,leftmargin=0pt,nosep]

            \item[\textbf{Cooperating detection service:}]
            We assume a cooperating AI vendor or enterprise-controlled monitoring proxy shared across vendors runs extraction and verification on inbound plaintext prior to model processing. The vendor requires only the ability to execute the extraction algorithm and perform key-based verification; it need not have prior access to the organization's files. Key distribution and notification protocols are considered important but out of scope.

            \item[\textbf{In-scope artifacts and channel:}]
            We restrict attention to text-based files and plaintext content submitted to cloud-hosted AI services (pasted into a chat interface or uploaded via API). Our evaluation targets transform suites representative of real-world text handling and sanitization.

            \item[\textbf{Local and offline LLMs:}]
            An adversary running a locally hosted LLM bypasses the vendor-side ingestion boundary entirely and falls outside this detection model. This is an acceptable scope restriction as the current dominant enterprise and adversarial deployment mode is cloud-hosted, agentic malware such as \textsc{Ransomware~3.0} deliberately targets external APIs to avoid staging large model weights on victim infrastructure, and local-LLM exfiltration/execution is more naturally addressed by on-host monitors and defenses. The framework remains composable with such controls in a zero-trust, defense-in-depth deployment and we revisit this paradigm in Section~\ref{sec:local-llm}.
    
            \item[\textbf{Out of scope:}]
            We explicitly exclude: (i)~binary and multimedia steganography; (ii)~file-format watermarks (PDF metadata, OOXML hidden fields) as a primary detection signal; (iii)~network-level exfiltration detection; (iv)~automated generation of realistic canary content at scale; and (v)~the privacy and legal framework for vendor cooperation. Furthermore, we treat Tier~4 transforms as an upper-bound adversarial capability evaluated for completeness rather than robustness.

        \end{description}

   \begin{figure}[t] 
        \centering
        \includegraphics[width=1\linewidth]{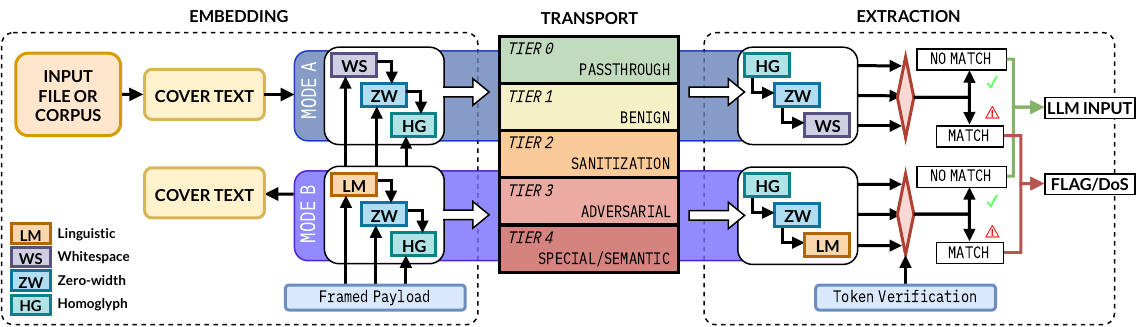}
        \caption{Framework pipeline overview including encoding stacks, possible transforms, and inverse decoding. Any verified recovery constitutes detection.}
        \label{fig:pipeline}
        \vspace{-1em}
    \end{figure}

\section{Framework Design \& Implementation} \label{sec:architecture}

    This section presents the steganographic canary framework end to end. We organize the presentation by following the data path: secret creation (Section~\ref{sec:secret-gen}), embedding (Section~\ref{sec:embedding-methods}), layering and orchestration (Section~\ref{sec:stacking}), and intended vendor-side detection (Section~\ref{sec:detection}). 

    \subsection{Pipeline Overview}

        Our framework targets the gap identified in our threat model where AI-powered malware and insider users submit sensitive plaintext to cloud-hosted LLMs through workflows that bypass traditional perimeter controls. Rather than attempting to prevent exfiltration at the network boundary, our architecture places detection at the AI service's processing boundary by embedding cryptographically verifiable identifiers into canary documents that survive common text transformations.
        Figure~\ref{fig:pipeline} summarizes this pipeline. The system generates a per-file identifier and embeds it into plaintext canaries,  enabling a cooperating AI service to extract and verify the identifier before LLM processing. The identifier is embedded across complementary in-text channels, allowing detection  if any layer survives transport. We construct and empirically verify two main modes of deployment:

        \begin{figure}[h]
        \centering
        \begin{minipage}[t]{0.47\textwidth}
            \centering
            {\footnotesize\textbf{Mode~A Encode} (Symbolic Stack)}\\[2pt]
            \(
                \textit{cover}
                \;\xrightarrow{\textsc{WS}.{\textsc{enc}}}\;
                \xrightarrow{\textsc{ZW}.{\textsc{enc}}}\;
                \xrightarrow{\textsc{HG}.{\textsc{enc}}}\;
                \textit{canary}
            \)
        \end{minipage}%
        \hfill
        \begin{minipage}[t]{0.5\textwidth}
            \centering
            {\footnotesize\textbf{Mode~B Encode} (Hybrid Stack)}\\[2pt]
            \(
                \textit{payload}
                \;\xrightarrow{\textsc{LM}.{\textsc{enc}}}\;
                \textit{cover}
                \xrightarrow{\textsc{ZW}.{\textsc{enc}}}\;
                \xrightarrow{\textsc{HG}.{\textsc{enc}}}\;
                \textit{canary}
            \)
        \end{minipage}
        \end{figure}

        \begin{itemize}[noitemsep,nolistsep,leftmargin=*]
            \item \textbf{Mode~A}\;--\;\textbf{Mark existing documents (symbolic stack):} Given an existing plaintext document, the framework applies a sequence of symbolic encoders that each embed the full secret independently on a disjoint character surface, enabling low-friction retrofitting of existing sensitive files without altering visible content.
        
            \item \textbf{Mode~B}\;--\;\textbf{Generate canary documents (hybrid stack):} The linguistic encoder generates synthetic text whose token choices encode the secret, on top of which compatible symbolic layers are then added as independent detection channels. WS artifacts can be deleted (rather than normalized) by common sanitization, thereby irreversibly corrupting the byte stream required for LM, and so we omit WS in this mode.
        \end{itemize}

        \noindent The framework is designed such that any stego-technique that conforms to the module interface defined in Section~\ref{sec:funcs} can be substituted for, or added alongside, the methods evaluated here. The methods, WS, ZW, HG, and LM, were selected to span the symbolic and linguistic method classes and to demonstrate the composition principles, however they are not the only viable implementations.

    \subsection{Secret Token Generation \& Framing} \label{sec:secret-gen}
    
        Each canary embeds a fixed-size, per-file identifier derived from organization-held key material. We validate the framework via two verification schemes that differ in key management but share the same framing and embedding pipeline, although other schemes may be easily implemented.
        
        \subsubsection{Shared-Key Verification (HMAC-SHA256):}
        Given a unique file identifier $\mathit{file\_id}$ (path, hash, or UUID) and an organization key $k_{\text{org}}$, we compute an HMAC-SHA256 tag and truncate it to the first 16~bytes, yielding a 128-bit token:
        
        \begin{center}
        \begin{minipage}[t]{0.48\textwidth}
            \centering
            {\footnotesize\textbf{Token Construction}}\\[2pt]
            \(
                \mathit{token} = \texttt{HMAC-SHA256}(k_{\text{org}},\;\mathit{file\_id})\,[{:}16]
            \)
        \end{minipage}%
        \hfill
        \begin{minipage}[t]{0.48\textwidth}
            \centering
            {\footnotesize\textbf{Framed payload} (18~bytes)}\\[2pt]
            \(
                \mathit{payload} =
                \underbrace{\texttt{len}(token)_{\text{BE16}}}_{2\;\text{bytes}}
                \;\|\;
                \underbrace{\mathit{token}}_{16\;\text{bytes}}
            \)
        \end{minipage}
        \end{center}
        
        \noindent The vendor verifies recovered payloads against a pre-loaded registry of valid tokens ($10^{3}$~organizations $\times$ $10^{4}$~file IDs $= 10^{7}$~entries in our evaluation). A random 128-bit candidate matches any registry entry with probability $P_{\mathrm{fp}} \leq 10^{7} \times 2^{-128} \approx 2.9 \times 10^{-32}$.
        
        \subsubsection{Public-Key Verification (Ed25519 via EdDSA):}
        The organization derives an Ed25519 key pair $(\mathit{sk}, \mathit{pk})$ deterministically from $k_{\text{org}}$ using the EdDSA (Edwards-curve Digital Signature Algorithm). For each file, a 4-byte identifier $\mathit{file\_uuid} = \texttt{SHA\text{-}256}(\mathit{file\_id})[{:}4]$ is signed:
        
        \begin{center}
        \begin{minipage}[t]{0.47\textwidth}
            \centering
            {\footnotesize\textbf{Token Construction}}\\[2pt]
            \(
                \mathit{sig} = \texttt{Ed25519\_Sign}(\mathit{sk},\;\mathit{file\_uuid})
            \)\\[2pt]
            \(
                \mathit{token} = \lbrack\mathit{file\_uuid}\;\|\;\mathit{sig}\rbrack
            \)
        \end{minipage}%
        \hfill
        \begin{minipage}[t]{0.52\textwidth}
            \centering
            {\footnotesize\textbf{Framed payload} (70~bytes)}\\[2pt]
            \(
                \mathit{payload} =
                \lbrack\underbrace{\texttt{len}(token)_{\text{BE16}}}_{2\;\text{bytes}}
                \;\|\;
                \underbrace{\mathit{file\_uuid}}_{4\;\text{bytes}}
                \;\|\;
                \underbrace{\mathit{sig}}_{64\;\text{bytes}}\rbrack
            \)
        \end{minipage}
        \end{center}
        
        \noindent The vendor stores only the organization's 32-byte public key $\mathit{pk}$ and verifies any recovered payload via $\texttt{Ed25519\_Verify}(\mathit{pk},\, \mathit{file\_uuid},\, \mathit{sig})$, providing ${\sim}$128-bit security without requiring a per-file token registry or shared-secret distribution~\cite{bernstein2006curve}. A forged signature verifies with probability ${\approx}\,2^{-128}$ per key, where with $10^{3}$~registered organizations, $P_{\mathrm{fp}} \leq 10^{3} \times 2^{-128} \approx 2.9 \times 10^{-36}$. We refer to this scheme as EdDSA in subsequent sections of the paper.
        
        Both schemes are deterministic and provide ${\sim}$128-bit security. HMAC produces an 18-byte framed payload and requires the vendor to maintain a shared key or token registry whereas EdDSA produces a 70-byte framed payload but requires only a one-time public-key registration eliminating registry synchronization. The larger EdDSA payload increases symbolic capacity requirements but remains within the capacity of typical prose (Section~\ref{sec:experimental_setup}).
        Before embedding, both schemes prepend a 2-byte big-endian length prefix to create a self-delimiting framed payload. For Mode~B, the linguistic encoder additionally pads the framed payload to ensure the generated cover text provides sufficient capacity for subsequent symbolic layers. This padding budget is adjusted per scheme to accommodate the difference in payload size for cascading embedding layers.

            \begin{figure}[t]
            \centering
            \footnotesize
            \begin{subfigure}[t]{0.48\linewidth}
                \hrule
                \vspace{2pt}
                \textbf{Encode}$(\mathit{text},\; \mathit{payload},\; \mathcal{A},\; b)$
                \begin{algorithmic}[1]
                    \State $D \gets \textsc{ToDigits}(\texttt{len}(\mathit{payload})_{\text{BE16}} \| \mathit{payload},\; b)$
                    \State $P \gets \textsc{FindEligible}(\mathit{text},\; \mathcal{A})$
                    \State \textbf{if} $|P| < |D|$ \textbf{then return} \texttt{None}
                    \State $S \gets \textsc{SelectPositions}(P,\; |D|)$
                    \State \textbf{for} $i \gets 0$ \textbf{to} $|D|-1$ \textbf{do}
                        $\mathit{text}[S[i]] \gets \mathcal{A}[D[i]]$
                    \State \Return $\mathit{text}$
                \end{algorithmic}
                \vspace{4pt}
                \textbf{Decode}$(\mathit{text},\; \mathcal{A},\; b)$
                \begin{algorithmic}[1]
                    \State $D \gets [\ ]$
                    \State \textbf{for} $c \in \mathit{text}$ \textbf{do if} $c \in \mathcal{A}$\textbf{:}
                        $D.\text{append}(\mathcal{A}^{-1}[c])$
                    \State $\mathit{raw} \gets \textsc{FromDigits}(D,\; b)$
                    \State $n \gets \texttt{int}(\mathit{raw}[{:}2])$
                    \State \Return $\mathit{raw}[2{:}2{+}n]$
                \end{algorithmic}
                \vspace{2pt}
                \hrule
                \caption{Generic symbolic encoding and decoding. WS and ZW instantiate with $b{=}4$; HG with $b{=}2$ (bit-level). \textsc{SelectPositions} distributes stride-interleaved for ZW, head-first for WS/HG.}
                \label{alg:symbolic-generic}
            \end{subfigure}
            \hfill
            \begin{subfigure}[t]{0.48\linewidth}
                \hrule
                \vspace{2pt}
                \textbf{Encode}$(\mathit{payload})$
                \begin{algorithmic}[1]
                    \State $\mathit{prefix} \gets \texttt{len}(\mathit{payload})_{\text{BE16}}$
                    \State $\mathit{msg} \gets \textsc{ToBits}(\textsc{Pad}(\mathit{prefix} \| \mathit{payload}))$
                    \State $\mathit{ctx} \gets \textsc{Tokenize}(\mathit{context\_str})$
                    \State $\mathit{tokens} \gets \textsc{SAAC\text{-}Enc}(\text{GPT-2},\;\mathit{msg},\;\mathit{ctx})$
                    \State \Return $\textsc{Canonicalize}(\textsc{Detokenize}(\mathit{tokens}))$
                \end{algorithmic}
                \vspace{4pt}
                \textbf{Decode}$(\mathit{text})$
                \begin{algorithmic}[1]
                    \State $\mathit{text} \gets \textsc{Canonicalize}(\mathit{text})$
                    \State $\mathit{ctx} \gets \textsc{Tokenize}(\mathit{context\_str})$
                    \State $\mathit{msg} \gets \textsc{SAAC\text{-}Dec}(\text{GPT-2},\;\mathit{text},\;\mathit{ctx})$
                    \State $\mathit{raw} \gets \textsc{FromBits}(\mathit{msg})$
                    \State $n \gets \texttt{int}(\mathit{raw}[{:}2])$
                    \State \Return $\mathit{raw}[2{:}2{+}n]$
                \end{algorithmic}
                \vspace{2pt}
                \hrule
                \caption{SAAC linguistic encoding/ decoding over GPT-2. \textsc{Pad} extends payload per token  to ensure cover-text length for downstream symbolic layers. \textsc{Canonicalize} strips symbolic encoding  before linguistic decode.}
                \label{alg:saac}
            \end{subfigure}
            \caption{Pseudocode for symbolic (left) and linguistic (right) encoding families. Both produce a self-delimiting framed payload recoverable without external metadata.}
            \label{fig:algorithms}
            \end{figure}
    
    \subsection{Embedding Methods} \label{sec:embedding-methods}

        We employ four embedding methods (see Table~\ref{tab:method-comparison}) spanning two classes that include three symbolic methods that modify text via disjoint Unicode character surfaces, and one linguistic method that generates synthetic cover text encoding the secret in token-level statistical properties. 

        \begin{table}[b]
            \centering\small
            \caption{Comparison of embedding methods. WS, ZW, and HG are symbolic (character-level); LM is linguistic (token-level). Disjoint encoding surfaces enable stacking.}
            \label{tab:method-comparison}
            \setlength\extrarowheight{-4pt}
            \resizebox{\linewidth}{!}{%
            \begin{tabular}{@{}lcccc@{}}
                \toprule
                & \textbf{WS} & \textbf{ZW} & \textbf{HG} & \textbf{LM} \\
                \midrule
                \textbf{Method class}
                    & Symbolic & Symbolic & Symbolic & Linguistic \\
                \textbf{Encoding surface}
                    & Space codepoints & Inter-character gaps & Confusable glyphs & Token sequence \\
                \textbf{Encoding rule}
                    & Base-4 substitution & Base-4 insertion & 1-bit substitution & Arithmetic coding \\
                \textbf{Typical capacity}
                    & ${\sim}$0.5\,b/space & ${\sim}$2\,b/char & ${\sim}$0.35\,b/char & ${\sim}$2.5--5\,b/token \\
                \textbf{Edits existing text?}
                    & Yes & Yes (additive) & Yes & No (generates) \\
                \textbf{Length-preserving?}
                    & Yes & No (+ZW chars) & Yes & N/A \\
                \textbf{LM required?}
                    & No & No & No & Yes (GPT-2) \\
                \textbf{Dominant failure mode}
                    & Normalization / collapse & Zero-width stripping & Confusable norm. & Semantic rewriting \\
                \textbf{Primary vuln.\ tier}
                    & Tier~2 & Tier~3 & Tier~3 & Tier~4 \\
                \bottomrule
            \end{tabular}
            }%
        \end{table}

        \subsubsection{Symbolic Methods}

            The three symbolic methods share a common structure as they define an alphabet of visually equivalent Unicode variants, convert the framed payload to a digit stream in the alphabet's base, and then distribute symbols across the target character surface.
            Table~\ref{tab:method-comparison} summarizes the key properties of each method; the paragraphs below focus on encoding mechanics and implementation choices. The algorithm for symbolic encode and decode is given in Figure~\ref{alg:symbolic-generic}; the three methods instantiate it with the alphabets and capacity expressions derived below.

            \paragraph{Whitespace Substitution \textbf{(WS)}}
            WS replaces selected ASCII spaces (\texttt{U+0020}) with one of four visually identical Unicode variants, implementing a base-4 alphabet.
            Each payload byte consumes four space positions (four base-4 digits, LSB-first), and encoding proceeds left-to-right across the first $4(|\textit{prefix}| + |\textit{payload}|)$ ASCII spaces. 
            In typical English prose (${\sim}$100 spaces per 500 words), this yields ${\sim}$23 payload bytes. 
            Decoding scans for characters in $\mathcal{A}_{\textsc{ws}}$, reconstructs the base-4 digit stream, and parses the length-prefixed payload.
            WS is fragile to NFKC normalization and whitespace collapsing back to ASCII. We adapt the implementation of this method from \textsc{Innamark}~\cite{hellmeier2025innamark}.

            \begin{center}
            \begin{minipage}[t]{0.48\textwidth}
                \centering
                {\footnotesize\textbf{WS alphabet} \quad $b = 4$}\\[2pt]
                \(
                    \mathcal{A}_{\textsc{ws}} = \{\texttt{U+2008},\;\texttt{U+2009},\;\texttt{U+202F},\;\texttt{U+205F}\}
                \)
            \end{minipage}%
            \hfill
            \begin{minipage}[t]{0.48\textwidth}
                \centering
                {\footnotesize\textbf{WS capacity}}\\[2pt]
                \(
                    C_{\textsc{ws}} = \left\lfloor N_{\text{spaces}} / 4 \right\rfloor - 2 \;\text{bytes}
                \)
            \end{minipage}
            \end{center}

            \paragraph{Zero-Width Character Insertion \textbf{(ZW)}}
            This method embeds data in inter-character gaps by inserting invisible Unicode format characters between adjacent visible (non-newline) characters, using a four-symbol alphabet. 
            We distribute insertions (instead of clustering them) via stride-based interleaving, where for $n$ insertions across $m$ candidate gaps, the $i$-th symbol is placed at gap position $\lfloor i \cdot m / n \rfloor$. 
            Decoding collects all $\mathcal{A}_{\textsc{zw}}$ characters in document order and reconstructs the payload.
            ZW survives normalization that preserves format characters but is destroyed by targeted zero-width stripping (Tier~3). Since ZW is purely additive, stripping ZW restores the exact original byte stream, satisfying invertibility for cross-class stacking (Section~\ref{sec:stacking}). We implement this method through extension of \texttt{zwsp-steg}~\cite{zwsp-steg} and \texttt{unicode\_steganography.js}~\cite{330k-steg}.

            \begin{center}
            \begin{minipage}[t]{0.48\textwidth}
                \centering
                {\footnotesize\textbf{ZW alphabet} \quad $b = 4$}\\[2pt]
                \(
                    \mathcal{A}_{\textsc{zw}} = \{\texttt{U+200B},\;\texttt{U+200C},\;\texttt{U+200D},\;\texttt{U+FEFF}\}
                \)
            \end{minipage}%
            \hfill
            \begin{minipage}[t]{0.48\textwidth}
                \centering
                {\footnotesize\textbf{ZW capacity}}\\[2pt]
                \(
                    C_{\textsc{zw}} = \left\lfloor (N_{\text{visible}} - 1) / 4 \right\rfloor - 2 \;\text{bytes}
                \)
            \end{minipage}
            \end{center}

            \begin{table}[b]
            \centering\small
            \caption{Homoglyph confusable pairs used for HG encoding. Latin characters (roman) and their Cyrillic confusables (italic) are visually pixel-identical in common system fonts; Unicode codepoints disambiguate.}
            \label{tab:homoglyph-pairs}
            \setlength{\tabcolsep}{3.2pt}
            \begin{minipage}[t]{0.4\linewidth}
                \centering
                {\footnotesize\textbf{Lowercase (7 pairs)}}\\[4pt]
                \resizebox{0.75\textwidth}{!}{
                \begin{tabular}{@{}l*{7}{c}@{}}
                    \toprule
                    \textbf{Latin}     & a & c & e & o & p & x & y \\
                    \midrule
                    \textbf{Cyrillic}  & {a} & {c} & {e} & {o}
                              & {p} & {x} & {y} \\
                    \textbf{Codepoint} \qquad
                        & \rotatebox{90}{\footnotesize 0430}
                        & \rotatebox{90}{\footnotesize 0441}
                        & \rotatebox{90}{\footnotesize 0435}
                        & \rotatebox{90}{\footnotesize 043E}
                        & \rotatebox{90}{\footnotesize 0440}
                        & \rotatebox{90}{\footnotesize 0445}
                        & \rotatebox{90}{\footnotesize 0443} \\
                    \bottomrule
                \end{tabular}}
            \end{minipage}%
            \hfill
            \begin{minipage}[t]{0.55\linewidth}
                \centering
                {\footnotesize\textbf{Uppercase (11 pairs)}}\\[4pt]
                \resizebox{0.76\textwidth}{!}{
                \begin{tabular}{@{}l*{11}{c}@{}}
                    \toprule
                    \textbf{Latin}     & A & B & C & E & H & K & M & O & P & T & X \\
                    \midrule
                    \textbf{Cyrillic}  & {A} & {B} & {C} & {E} & {H}
                    & {K} & {M} & {O} & {P} & {T} & {X} \\
                    \textbf{Codepoint} \qquad
                        & \rotatebox{90}{\footnotesize 0410}
                        & \rotatebox{90}{\footnotesize 0412}
                        & \rotatebox{90}{\footnotesize 0421}
                        & \rotatebox{90}{\footnotesize 0415}
                        & \rotatebox{90}{\footnotesize 041D}
                        & \rotatebox{90}{\footnotesize 041A}
                        & \rotatebox{90}{\footnotesize 041C}
                        & \rotatebox{90}{\footnotesize 041E}
                        & \rotatebox{90}{\footnotesize 0420}
                        & \rotatebox{90}{\footnotesize 0422}
                        & \rotatebox{90}{\footnotesize 0425} \\
                    \bottomrule
                \end{tabular}}
            \end{minipage}
            \end{table}

            \paragraph{Homoglyph Substitution \textbf{(HG)}}
    
            encodes a bitstream by replacing Latin characters with visually indistinguishable Cyrillic confusables from Unicode UTS\,\#39~\cite{unicode2025uts39}.
            Each eligible position is a 1-bit channel, where the original Latin form denotes~0, and its Cyrillic counterpart denotes~1. In typical English prose, 35--40\% of characters are eligible.
            We restrict encoding to 18 high-confidence Latin--Cyrillic pairs (Table~\ref{tab:homoglyph-pairs}) chosen for pixel-level similarity under common fonts.
            To avoid ambiguity from pre-existing non-Latin characters, the encoder first normalizes the cover by mapping any Cyrillic confusables back to their Latin equivalents, then applies substitutions left-to-right according to the payload bitstream.
            HG is robust to transforms that preserve codepoint identity but fails under confusable normalization or ASCII transliteration (Tier~3). HG stripping maps Cyrillic substitutions back to their Latin originals satisfying byte-exact invertibility for hybrid stacking. We base the implementation on Rizzo et al.~\cite{rizzo2016} with confusable pairs from UTS\,\#39~\cite{unicode2025uts39}.

        \subsubsection{Linguistic Method (LM)} \label{sec:linguistic-method}
            
            This method encodes the secret in output token choices by generating synthetic cover text via Self-Adjusting Arithmetic Coding (SAAC)~\cite{shen2020saac} over GPT-2 (124\,M parameters)~\cite{radford2019gpt2}; the algorithm is given in Figure~\ref{alg:saac}. The framed payload is interpreted as a fractional value in $[0,1)$, and at each generation step SAAC partitions the next-token probability distribution and selects the token whose interval contains the current value, embedding information while producing fluent prose. SAAC dynamically adapts its truncation parameter~$K$ to per-step entropy and we set the remaining hyperparameters, context strings, and seeds to fixed values.
            %
            %
            Decoding replays the token sequence through the same model, reconstructs arithmetic-coding intervals, and recovers the embedded bitstream. Correctness requires bit-exact agreement on model weights, tokenizer, context string, and all SAAC parameters.

            The effective embedding rate is ${\sim}$2.5--5 bits per token (${\sim}$10 characters per payload byte). The SAAC token budget and minimum output length are adjusted per verification scheme to accommodate the respective framed payload size (18~bytes for HMAC, 70~bytes for EdDSA), with additional padding to ensure downstream symbolic layers have sufficient encoding surface. At model load time, the wrapper scans GPT-2's full 50{,}257-token vocabulary and suppresses any token whose decoded form would pollute downstream symbolic encoding surfaces. Because LM's signal resides in the token sequence, any byte-stream modification can cause decoding failure, and only symbolic layers whose stripping is byte-exact-invertible (ZW and HG) can be safely layered on top of the LM output. WS does not satisfy this requirement, as transport transforms can delete substituted Unicode spaces rather than normalizing them back to ASCII, irreversibly corrupting the byte stream. We validate this interaction empirically in Section~\ref{sec:evaluation}.
            We implement SAAC via \textsc{StegaText}~\cite{shen2020saac}, building on Ziegler et al.~\cite{ziegler2019neural}, with compatibility wrappers for newer \texttt{transformers}~4.x APIs.

            \begin{figure}[t]
            \centering
            \newcommand{\sig}[3]{%
                \textcolor{funcblue}{\texttt{\textbf{#1}}}%
                \texttt{(}\textcolor{argbrown}{\texttt{#2}}\texttt{)}%
                \;\textcolor{retgray}{\texttt{$\to$\,#3}}%
            }
            \begin{minipage}[t]{0.48\linewidth}
                \sig{encode}{text, payload}{text | None}\\[2pt]
                \footnotesize Embeds payload into text, or generates text
                carrying the payload (LM mode). Returns \texttt{None}
                if capacity is insufficient.
            \end{minipage}%
            \hfill
            \begin{minipage}[t]{0.48\linewidth}
                \sig{decode}{text}{payload | None}\\[2pt]
                \footnotesize Scans for embedded symbols and reconstructs
                a candidate payload. Returns \texttt{None} if no valid
                payload is found.
            \end{minipage}
            
            \vspace{0.7em}
            
            \begin{minipage}[t]{0.48\linewidth}
                \sig{strip\_encoding}{text}{text}\\[2pt]
                \footnotesize Removes method's artifacts and restores
                a byte-exact original, enabling downstream decoders in
                cross-class stacking.
            \end{minipage}%
            \hfill
            \begin{minipage}[t]{0.48\linewidth}
                \sig{capacity}{text}{int}\\[2pt]
                \footnotesize Returns the maximum number of payload bytes
                embeddable in the given text under this method's surface
                constraints.
            \end{minipage}
            \caption{Uniform function interface implemented by all embedding methods, characterized by function name, input arguments, and return objects.}
            \label{fig:module-interface}
            \end{figure}

    \subsection{Layering, Orchestration, and Transport Simulation} \label{sec:stacking}

        We compose the framework's embedding methods into configurations that provide layered (stacked) detection coverage. This subsection defines the module interface, the seven (individual and stacked) configurations used throughout the evaluation, and the encode/decode ordering that enables hybrid layering for implementation of the pipeline shown in Figure~\ref{fig:pipeline}.

        \subsubsection{Module Interface} \label{sec:funcs}

            Each embedding method conforms to the uniform interface shown in Figure~\ref{fig:module-interface}. 
            Other symbolic methods targeting different surfaces (e.g., variation selectors, combining characters) or linguistic methods using different models or coding schemes can be integrated without changes to the orchestration, stacking, or detection logic, as long as byte-exact reversal is implemented via \texttt{strip\_encoding()}.

            \begin{table}[t]
                \centering\small
                \caption{Method configurations. M1--M4 are individual baselines; M5 is the symbolic-only stack (Mode~A); M6 is the recommended hybrid stack (Mode~B); M7 includes WS and demonstrates cross-layer interference.}
                \label{tab:configurations}
                \resizebox{0.65\textwidth}{!}{
                \setlength\extrarowheight{-4pt}
                \begin{tabular}{@{}clcl@{}}
                \toprule
                \textbf{ID} & \textbf{Methods} & \textbf{Mode} & \textbf{Purpose} \\
                \midrule
                M1 & WS only             & - & Individual method baseline \\
                M2 & ZW only            & - & Individual method baseline \\
                M3 & HG only             & - & Individual method baseline \\
                M4 & LM only             & - & Individual method baseline \\
                \midrule
                M5 & WS + ZW + HG       & A & Symbolic defense-in-depth \\
                M6 & ZW + HG + LM       & B & Safe cross-class hybrid stack     \\
                M7 & WS + ZW + HG + LM  & - & Hybrid stack with interference \\
                \bottomrule
                \end{tabular}
                }
            \end{table}

        \subsubsection{Method Configurations and Stacking Semantics}
            
            We define seven embedding configurations (M1--M7 in Table~\ref{tab:configurations}) spanning single-method baselines, symbolic-only stacks, and hybrid layering. 
            M5 stacks the three symbolic methods on disjoint surfaces, providing redundancy against Tier~1--2 normalization. M6 is the recommended cross-class configuration as it applies ZW and HG over linguistically generated text, excluding WS because of the invertibility constraint discussed in Section~\ref{sec:linguistic-method}. M7 adds WS to the full stack to empirically expose this cross-layer interference.
            
            For Mode~A, the default order is WS~$\to$~ZW~$\to$~HG, and since the symbolic channels occupy disjoint surfaces, order does not affect correctness. For Mode~B, the linguistic encoder runs first to generate the cover text, after which compatible symbolic layers are applied. Decoding proceeds in reverse order, with each decoder first extracting a candidate payload and then calling \texttt{strip\_encoding} to remove its artifacts before passing restored text to the next stage:
            
            \begin{figure}[h]
            \centering
            \begin{minipage}[t]{0.46\textwidth}
                \centering
                {\footnotesize\textbf{Mode~A Decode} (Symbolic Stack)}\\[2pt]
                \(
                    \textit{canary}
                    \;\xrightarrow{\textsc{HG}.{\textsc{dec}}}\;
                    \xrightarrow{\textsc{ZW}.{\textsc{dec}}}\;
                    \xrightarrow{\textsc{WS}.{\textsc{dec}}}\;
                    \textit{payload}
                \)
            \end{minipage}%
            \hfill
            \begin{minipage}[t]{0.52\textwidth}
                \centering
                {\footnotesize\textbf{Mode~B Decode} (Hybrid Stack)}\\[2pt]
                \(
                    \textit{canary}
                    \;\xrightarrow{\textsc{HG}.{\textsc{dec}}}\;
                    \xrightarrow{\textsc{ZW}.{\textsc{dec}}}\;
                    \textit{cover}
                    \;\xrightarrow{\textsc{LM}.{\textsc{dec}}}\;
                    \textit{payload}
                \)
            \end{minipage}
            \end{figure}

        \subsubsection{Transport Simulation} \label{sec:transport-sim}

            To evaluate robustness under realistic handling and adversarial sanitization, we implement a transport simulator as a library of deterministic text transforms organized into the four tiers of the transport-transform taxonomy defined in the threat model (Section~\ref{sec:taxonomy}, Tables~\ref{tab:taxonomy}--\ref{tab:chains}). Each transform is a pure function ($f(\text{text}) \to \text{text}$), and composite transport chains are specified and executed as pre-ordered lists.

    \subsection{Vendor-Side Detection} \label{sec:detection}
        
\begin{figure}[t]
            \centering
            \includegraphics[width=\linewidth]{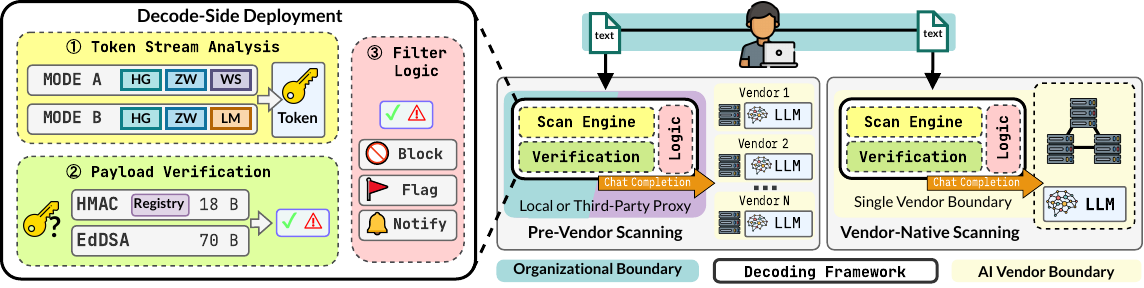}
            \caption{Detection via (a) vendor-native middleware or (b) local/third-party reverse proxy. The proxy/middleware intercepts outbound LLM API requests and  runs the decoding engine, verifying candidate payloads against the canary manifest/registry. A registry match triggers a configurable response logic before reaching the LLM.}
            \label{fig:vendor-detect}
        \end{figure}

        The detection component operates at the AI service's ingestion boundary where a pre-ingestion filter intercepts inbound plaintext before it reaches the LLM, runs extraction across all configured decoders, and verifies any candidate payloads. Under the HMAC scheme, verification is a hash-table lookup against a pre-loaded token registry. Under the EdDSA scheme, verification is a signature check against registered public keys. Detection is declared under the any-layer-recovers policy and the filter can be deployed in two configurations as outlined in Figure~\ref{fig:vendor-detect}.

        \begin{itemize}[noitemsep,nolistsep,leftmargin=*]
            \item \textbf{Reverse proxy} interposed between the client and the LLM API endpoint, extracting plaintext from requests and decoding in realtime pre-LLM invocation.
            \item \textbf{Vendor-native middleware} integrated into the AI service's request-processing pipeline, operating on parsed input text alongside other input classifiers pre-LLM invocation.
        \end{itemize}

        \noindent For our experiments, we implement and validate the reverse-proxy configuration (shown in Figure~\ref{fig:rware_reverse_proxy}) as an asynchronous HTTP service that sits between an OpenAI-compatible client and a locally hosted LLM server (\textsc{Ollama}). On each inbound chat-completion request, the proxy extracts message content and feeds it to a multi-layer scan engine that deploys in two phases:
        
        \begin{enumerate}[noitemsep,nolistsep,leftmargin=*]
            \item \textbf{Symbolic extraction.} For each registered symbolic decoder (WS, ZW, HG), the engine attempts \texttt{decode()}; if a payload is returned, it is checked against the manifest of known hashes. Regardless of match, the engine can call \texttt{strip\_encoding()} on the text and recurse to extract deeper layers, up to a configurable depth limit.
            \item \textbf{Sequential decoding.} After symbolic extraction, the scan engine runs the linguistic decoder on the cleaned text. Because SAAC requires bit-exact token agreement, symbolic artifacts must first be removed. In practice, however, full decoding is not always necessary as the architecture declares detection as soon as the first recovered payload is verified.
        \end{enumerate}
        
        \noindent Extracted payloads are verified according to the configured scheme. Under HMAC, the proxy checks against a pre-loaded manifest of known tokens keyed by hex digest, maintained as a local hash table for single-organization deployments or queried via a verification API for multi-tenant setups. Under EdDSA, the proxy verifies the embedded signature against the organization's registered public key, eliminating per-file registry updates and shared-secret distribution. Verified matches trigger a configurable response, demonstrated by a lockdown state in our proxy implementation that rejects all subsequent LLM requests.
        The decoder requires no access to the organization's files, no knowledge of which documents are canaries, and no semantic understanding of the content. The filter adds latency only to the extraction and verification steps (quantified in Section~\ref{sec:evaluation}).

\section{Experimental Setup} \label{sec:experimental_setup}

    This section defines our evaluation methodology, including experimental structure, test corpus, environment, and evaluation metrics. Technical details of the framework components, method configurations, and transport transforms are defined in Section~\ref{sec:architecture} and Section~\ref{sec:threat_model}, respectively.

    \subsection{Overview}
        We structure the evaluation as five controlled experiments and one case study, where Table~\ref{tab:experiment-grids} summarizes the dimensions of each and justification of each experiment follows. Embedding configurations are defined in Table~\ref{tab:configurations} and transport transforms and composite chains in Tables~\ref{tab:taxonomy}--\ref{tab:chains}.

        \begin{itemize}[noitemsep,nolistsep,leftmargin=28pt]
            \item[(\textbf{\S\ref{sec:results_exp1}})] \textbf{Baseline Robustness} localizes each failure surface across all individual transforms and 100 input files, where symbolic methods use corpus prose and LM uses generated cover texts.
            
            \item[(\textbf{\S\ref{sec:results_exp2a}})] \textbf{Stack Compatibility and Feasibility} verifies that all methods can coexist in a full stack at Tier-0, validating inverse-order decoding and byte-exact restoration before any transport.
            
            \item[(\textbf{\S\ref{sec:results_exp2b}})] \textbf{Layered Stacking Robustness} evaluates the defense-in-depth paradigm, reporting both per-layer and ANY-layer recovery across six composite chains.
            
            \item[(\textbf{\S\ref{sec:results_exp3}})] \textbf{False-Positive Rejection} runs each decoder on 100 unmarked documents from the corpus (Section~\ref{sec:corpus}) and verifies output under both verification schemes: HMAC against a $10^{7}$-token registry ($10^{3}$~organization keys $\times$ $10^{4}$~file IDs), and EdDSA against $10^{3}$~registered public keys.
    
    \item[(\textbf{\S\ref{sec:results_exp4}})] \textbf{Overhead and Timing} measures encoding/decoding latency across 100 files per mode under both verification schemes.
            
    \item[(\textbf{\S\ref{sec:results_e2e}})] \textbf{End-to-End Ransomware Case Study} simulates the exfiltration scenario from the threat model, exercising the canary lifecycle from seeding to vendor-side detection.
        \end{itemize}

        \begin{table}[h]
            \centering\small
            \caption{Per-experiment grid dimensions.}
            \label{tab:experiment-grids}
            \setlength\extrarowheight{-4pt}
            \resizebox{0.75\textwidth}{!}{
            \begin{tabular}{@{}clcc@{}}
                \toprule
                \textbf{Section} & \textbf{Experiment} & \textbf{Methods} & \textbf{Transports} \\
                \midrule
                (\S\ref{sec:results_exp1})
                    & Baseline Robustness Per Method
                    & 4 (M1--M4) & 13 transforms (Table~\ref{tab:taxonomy}) \\              
                (\S\ref{sec:results_exp2a})
                    & Stack Compatibility \& Feasibility
                    & 1 (M7)     & Tier-0  \\
                (\S\ref{sec:results_exp2b})
                    & Layered Stacking Robustness
                    & 7 (M1--M7) & 6 chains (Table~\ref{tab:chains}) \\
                \midrule
                (\S\ref{sec:results_exp3})
                    & False-Positive Rejection
                    & 4 decoders & Tier-0  \\
                (\S\ref{sec:results_exp4})
                    & Overhead \& Timing
                    & 7 (M1--M7) & Tier-0 \\
                \midrule
                (\S\ref{sec:results_e2e})
                    & End-to-End Ransomware Case Study
                    & 2 (M5, M6) & Tier-0  \\
                \bottomrule
            \end{tabular}
            }
        \end{table}
    
    \subsection{Test Corpus \& Token Scheme Assignment} \label{sec:corpus}
        
        We use a test corpus consisting of 100 English prose documents pulled from Wikipedia via the HuggingFace \texttt{datasets} library, filtered to a minimum of 3{,}000~characters and truncated to that target at word boundaries, yielding ${\sim}$3\,KB documents~\cite{wikimedia2023wikipedia}. Linguistically generated canaries range from ${\sim}$1.2\,KB (HMAC scheme) to ${\sim}$2\,KB (EdDSA) depending on payload size, while symbolic-only encodings preserve the original document size. 
        Each file is assigned identifiers under both the HMAC and EdDSA schemes per Section~\ref{sec:secret-gen} using a fixed evaluation key $k_{\text{org}}$. For experiments, symbolic configurations use $\mathit{file\_id} = \texttt{`dir\_name/\textit{file\_name}'}$ and linguistic configurations use $\mathit{file\_id} = \texttt{`generated\_\textit{id}'}$. For configurations including the linguistic encoder, SAAC generates its own cover text per Section~\ref{sec:linguistic-method}. Robustness experiments (Sections~\ref{sec:results_exp1}--\ref{sec:results_exp2b}) are presented using the HMAC scheme since transport-transform recovery depends on encoding-surface survival, not payload size. Deployment experiments (Sections~\ref{sec:results_exp3}--\ref{sec:results_exp4}) are run under both schemes to characterize the verification and timing trade-offs.

    \subsection{Environment and Setup} \label{sec:evironment}
    
        All experiments were conducted on a workstation with an Intel Core i9-12900K CPU, 96\,GB DDR5 RAM, and an NVIDIA RTX PRO 4000 GPU. All language models and latency measurements were executed locally on this hardware. The end-to-end ransomware case study uses an isolated virtual environment, with LLM-driven components served on the local network via the reverse-proxy described in Section~\ref{sec:detection}, interfaced with \textsc{Ollama}. This configuration is depicted in Figure~\ref{fig:vendor-detect} in Section~\ref{sec:results_exp4}.
        The framework is implemented in Python~3.11+ as a set of pluggable method modules conforming to the interface in Figure~\ref{fig:module-interface}. All symbolic methods and transport transforms require only the Python standard library (\texttt{re}, \texttt{unicodedata}); the linguistic method additionally requires PyTorch~2.x and HuggingFace \texttt{transformers}~4.x.
        Random seeds were fixed for both Python \texttt{random} and \texttt{torch.manual\_seed()} within the experimental runner. For timing measurements, we instrument encode and decode stages using \texttt{time.perf\_counter()}. 
        We conduct our end-to-end ransomware case study within the \textsc{SaMOSA} sandbox~\cite{udeshi2025samosa}, which provides time-synchronized side-channel telemetry and FakeNet network emulation for safe Linux malware execution.

        For experimental reproducibility, we fix all random seeds, locally cache the GPT-2 Model and BPE tokenizer, and isolate dependency version drift for \textsc{StegaText} via compatibility wrappers  introduced in Section~\ref{sec:linguistic-method}. A tokenizer shim is used here to expose the legacy \texttt{.encoder}/\texttt{.decoder} vocabulary dictionaries, from which a model wrapper then adjusts the return signature and suppresses tokens whose decoded form is not invariant under canonicalization, and a cache-management patch handles the \texttt{DynamicCache} object introduced in \texttt{transformers}~4.36. Together, these ensure that the same canary remains decodable across toolchain updates. We implement transform Tiers~1--3 using only standard Python library components (\texttt{unicodedata}, \texttt{re}). Tier~4 (LLM paraphrase) is implemented via a locally hosted LLM (\textsc{Ollama}) using the same model deployment as the end-to-end case study (Section~\ref{sec:results_e2e}), providing genuine semantic rewriting via an open-weights LLM. We execute all experiments via a single CLI entry point and serialize results to CSV.
    
    \subsection{Evaluation Metrics} \label{sec:eval-metrics}
        
        \begin{itemize}[noitemsep,nolistsep,leftmargin=*]
            \item \textbf{Recovery Rate (RR):}
            Defined as the fraction of trials that recover the embedded identifier exactly. For stacked configurations (M5--M7), we evaluate detection success using RR per-layer, indicating whether each individual encoding surface survived transport, and using an \textbf{ANY} (union), where a trial succeeds if at least one layer recovers the correct token. 
        
            \item \textbf{Encoding/Decoding Time ($T_{\text{enc}}$, $T_{\text{dec}}$):}
            This is the wall-clock time for the full encode or decode stage (milliseconds for CPU and seconds for GPU), including all layers for stacked configurations. 
        \end{itemize}
        
        \noindent 
        Experiments show that transport transforms either preserve the encoding surface entirely (RR${=}100\%$) or destroy it completely (RR${=}0\%$) for any given file, with no partial-corruption regime. Thus continuous metrics such as bit error rate (BER) and capacity utilization provide no additional information beyond RR and are omitted, along with error correction schemes. Population-level survival rates below 100\% (e.g., 98\% for LM under T10) reflect the fraction of files whose encoding surface survived, \textbf{not partial recovery} within any single file.
        The linguistic method exhibits a minor deviation from this per-file binary pattern where GPT-2's byte-level BPE vocabulary occasionally produces tokens containing non-ASCII Unicode characters that survive encoding but interfere arithmetic-coding retokenization during decoding. We analyze this artifact further in Section~\ref{sec:discussion}. 
        
\section{Evaluation Results} \label{sec:evaluation}
    
    \subsection{Per-Method Robustness Baselines} \label{sec:results_exp1}

        We evaluate each individual method (M1--M4) under a Tier-0 baseline and all 12 individual transforms (T01--T12), across 100 files per method. Table~\ref{tab:exp1-heatmap} presents the per-transform ablation.
        Results show that all four methods achieve 100\% recovery at Tier~0, confirming baseline feasibility for embedding and recovering a per-file cryptographic identifier from realistic plaintext under ideal conditions.\footnote{Robustness results are shown for the HMAC scheme. EdDSA results were found to be experimentally equivalent as recovery depends on encoding-surface survival, not payload size, providing no new information beyond the RR presented here.} They remain intact through Tier~1 (T01--T04), showing that copy-paste, line reflow, smart-quote replacement, and trailing-whitespace stripping does not threaten recovery.

        \begin{table}[h]
            \centering\small
            \caption{Per-transform ablation for baseline robustness. Recovery rate (RR) for each individual method under Tier-0 and T01--T12. Each cell aggregates 100 files. Color indicates recovery rate: \colorbox{heatHigh}{100\%}, \colorbox{heatNear}{98\%}, \colorbox{heatLow}{8\%}, \colorbox{heatZero}{0\%}.}
            \label{tab:exp1-heatmap}
            \resizebox{0.65\textwidth}{!}{
            \begin{tabular}{@{}llcccc@{}}
                \toprule
                & \textbf{Transform}
                    & \textbf{M1 (WS)} & \textbf{M2 (ZW)}
                    & \textbf{M3 (HG)} & \textbf{M4 (LM)} \\
                \midrule
                \multirow{1}{*}{\rotatebox{90}{\scriptsize T.0}}
                & Tier-0 (none)
                    & \HH{100}{100} & \HH{100}{100}
                    & \HH{100}{100} & \HH{100}{100} \\
                \midrule
                \multirow{4}{*}{\rotatebox{90}{\scriptsize Tier 1}}
                & T01 -- copy/paste
                    & \HH{100}{100} & \HH{100}{100}
                    & \HH{100}{100} & \HH{100}{100} \\
                & T02 -- line reflow
                    & \HH{100}{100} & \HH{100}{100}
                    & \HH{100}{100} & \HH{100}{100} \\
                & T03 -- smart quotes
                    & \HH{100}{100} & \HH{100}{100}
                    & \HH{100}{100} & \HH{100}{100} \\
                & T04 -- trailing WS strip
                    & \HH{100}{100} & \HH{100}{100}
                    & \HH{100}{100} & \HH{100}{100} \\
                \midrule
                \multirow{3}{*}{\rotatebox{90}{\scriptsize Tier 2}}
                & T05 -- NFKC normalize
                    & \HH{0}{0}     & \HH{100}{100}
                    & \HH{100}{100} & \HH{100}{100} \\
                & T06 -- collapse WS
                    & \HH{0}{0}     & \HH{100}{100}
                    & \HH{100}{100} & \HH{100}{100} \\
                & T07 -- strip format chars
                    & \HH{100}{100} & \HH{0}{0}
                    & \HH{100}{100} & \HH{100}{100} \\
                \midrule
                \multirow{3}{*}{\rotatebox{90}{\scriptsize Tier 3}}
                & T08 -- strip ZW
                    & \HH{100}{100} & \HH{0}{0}
                    & \HH{100}{100} & \HH{100}{100} \\
                & T09 -- homoglyph norm.
                    & \HH{100}{100} & \HH{100}{100}
                    & \HH{0}{0}     & \HH{100}{100} \\
                & T10 -- full ASCII strip
                    & \HH{0}{0}     & \HH{0}{0}
                    & \HH{0}{0}     & \HH{98}{98} \\
                \midrule
                \multirow{2}{*}{\rotatebox{90}{\scriptsize Tier 4}}
                & T11 -- punct./case strip
                    & \HH{100}{100} & \HH{100}{100}
                    & \HH{8}{8}     & \HH{0}{0} \\
                & T12 -- LLM paraphrase
                    & \HH{0}{0}     & \HH{0}{0}
                    & \HH{0}{0}     & \HH{0}{0} \\
                \bottomrule
            \end{tabular}
            }
        \end{table}

        We observe divergence at Tier~2, exposing largely orthogonal failure surfaces. WS fails under NFKC normalization (T05) and whitespace collapsing (T06) as expected, while ZW fails under format-character stripping (T07) because its alphabet lies in Unicode category~Cf, unlike WS characters in category~Zs. HG and LM survive all Tier~2 transforms. At Tier~3, each targeted transform destroys its intended symbolic method, as T08 breaks ZW, T09 breaks HG, and T10 removes all symbolic channels. LM remains through T10 with 98\% recovery, consistent with the GPT-2 vocabulary artifact discussed in Section~\ref{sec:eval-metrics}. Tier~4 reveals an inverse pattern across methods as T11 preserves WS and ZW but destroys LM and nearly destroys HG (8\%, due to cover text containing no capital letters), while T12 destroys all methods. 

    \subsection{Defense-in-Depth via Layered Composition} \label{sec:results_stacking}
    
        \subsubsection{Stacking Feasibility and Compatibility} \label{sec:results_exp2a}

            We establish Tier-0 compatibility using 100 generated texts for the full stack (M7: WS+ZW+HG+LM) in Figure~\ref{fig:exp2a-compat}, showing all four layers coexist, and inverse-order decoding recovers the embedded payloads correctly. 
            The control column shows that the linguistic encoder achieves 100\% recovery before symbolic layering. After full-stack encoding, all four layers recover 98 of 100 tokens, and the Restored column shows that symbolic stripping reconstructs the original linguistic output byte-for-byte in those same 98 cases. The remaining 2\% loss is due to the GPT-2 vocabulary artifact, where non-ASCII tokens in 2 of the 100 texts disrupt symbolic decoding. This validates M7, and therefore M6 (its subset), under ideal conditions, showing that remaining interference is transport-induced rather than introduced by stacking itself.

            \tcbset{statcard/.style={
                colback=white,
                colframe=black!40,
                boxrule=0.4pt,
                arc=1pt,
                left=2pt, right=2pt,
                top=2pt, bottom=2pt,
                boxsep=0pt,
                before skip=0pt, after skip=0pt
            }}
            
            \begin{figure}[h]
            \centering
            \newcommand{\statbox}[2]{%
                \begin{minipage}[t]{0.15\linewidth}
                    \begin{tcolorbox}[statcard]
                        \centering
                        \scriptsize\textbf{#1}\\[1pt]
                        \normalsize\textbf{#2}
                    \end{tcolorbox}
                \end{minipage}%
            }
            \statbox{Control\\Decode}{100/100}\hspace{0.01\linewidth}%
            \statbox{Whitespace\\Subst.}{98/100}\hspace{0.01\linewidth}%
            \statbox{Zero-Width\\Insert}{98/100}\hspace{0.01\linewidth}%
            \statbox{Homoglyph\\Subst.}{98/100}\hspace{0.01\linewidth}%
            \statbox{Linguistic\\Model}{98/100}\hspace{0.01\linewidth}%
            \statbox{Byte-Exact\\Restored}{98/100}
            \caption{Stack compatibility (M7, Tier-0, 100\,files). Control decode verifies linguistic encoding before symbolic layers are applied, and  byte-exact Restored confirms byte-for-byte recovery after inverse-order extraction.}
            \label{fig:exp2a-compat}
            \end{figure}

        \subsubsection{Layered Stacking Robustness} \label{sec:results_exp2b}

            \begin{table*}[t]
                \centering\small
                \caption{Layered configuration comparison (7 configs $\times$ 6 composite chains $\times$ 100 files). RR = recovery rate. For M5--M7, \textbf{ANY} = union (at least one layer recovers); indented rows show per-layer survival. 
                }
                \label{tab:exp2b-layered}
                \resizebox{0.8\textwidth}{!}{
                \begin{tabular}{@{}lcccccc@{}}
                    \toprule
                    \textbf{Configuration}
                        & \textbf{Tier-1} & \textbf{Tier-2} & \textbf{Tier-3}
                        & \textbf{Tier-1+2} & \textbf{Tier-1+2+3} & \textbf{Tier-4} \\
                    \midrule
                    \textbf{M1 (WS)}
                        & \HHH{100}{100} & \HHH{0}{0}   & \HHH{0}{0}
                        & \HHH{0}{0}     & \HHH{0}{0}   & \HHH{0}{0} \\
                    \textbf{M2 (ZW)}
                        & \HHH{100}{100} & \HHH{0}{0}   & \HHH{0}{0}
                        & \HHH{0}{0}     & \HHH{0}{0}   & \HHH{0}{0} \\
                    \textbf{M3 (HG)}
                        & \HHH{100}{100} & \HHH{100}{100} & \HHH{0}{0}
                        & \HHH{100}{100} & \HHH{0}{0}     & \HHH{0}{0} \\
                    \textbf{M4 (LM)}
                        & \HHH{100}{100} & \HHH{100}{100} & \HHH{98}{98}
                        & \HHH{100}{100} & \HHH{98}{98}   & \HHH{0}{0} \\
                    \midrule
                    \textbf{M5 (WS+ZW+HG)} (ANY)
                        & \HHH{100}{100} & \HHH{100}{100} & \HHH{0}{0}
                        & \HHH{100}{100} & \HHH{0}{0}     & \HHH{0}{0} \\
                    \quad WS layer
                        & \HHH{100}{100} & \HHH{0}{0}   & \HHH{0}{0}
                        & \HHH{0}{0}     & \HHH{0}{0}   & \HHH{0}{0} \\
                    \quad ZW layer
                        & \HHH{100}{100} & \HHH{0}{0}   & \HHH{0}{0}
                        & \HHH{0}{0}     & \HHH{0}{0}   & \HHH{0}{0} \\
                    \quad HG layer
                        & \HHH{100}{100} & \HHH{100}{100} & \HHH{0}{0}
                        & \HHH{100}{100} & \HHH{0}{0}     & \HHH{0}{0} \\
                    \midrule
                    \textbf{M6 (ZW+HG+LM)} (ANY)
                        & \HHH{99}{99} & \HHH{99}{99} & \HHH{97}{97}
                        & \HHH{99}{99} & \HHH{97}{97} & \HHH{0}{0} \\
                    \quad LM layer
                        & \HHH{99}{99} & \HHH{99}{99} & \HHH{97}{97}
                        & \HHH{99}{99} & \HHH{97}{97} & \HHH{0}{0} \\
                    \quad ZW layer
                        & \HHH{99}{99} & \HHH{0}{0}   & \HHH{0}{0}
                        & \HHH{0}{0}   & \HHH{0}{0}   & \HHH{0}{0} \\
                    \quad HG layer
                        & \HHH{99}{99} & \HHH{99}{99} & \HHH{0}{0}
                        & \HHH{99}{99} & \HHH{0}{0}   & \HHH{0}{0} \\
                    \midrule
                    \textbf{M7 (WS+ZW+HG+LM)} (ANY)
                        & \HHH{98}{98} & \HHH{98}{98} & \HHH{0}{0}
                        & \HHH{98}{98} & \HHH{96}{96} & \HHH{0}{0} \\
                    \quad LM layer
                        & \HHH{98}{98} & \HHH{98}{98} & \HHH{0}{0}
                        & \HHH{98}{98} & \HHH{96}{96} & \HHH{0}{0} \\
                    \quad WS layer
                        & \HHH{98}{98} & \HHH{0}{0}   & \HHH{0}{0}
                        & \HHH{0}{0}   & \HHH{0}{0}   & \HHH{0}{0} \\
                    \quad ZW layer
                        & \HHH{98}{98} & \HHH{0}{0}   & \HHH{0}{0}
                        & \HHH{0}{0}   & \HHH{0}{0}   & \HHH{0}{0} \\
                    \quad HG layer
                        & \HHH{98}{98} & \HHH{98}{98} & \HHH{0}{0}
                        & \HHH{98}{98} & \HHH{0}{0}   & \HHH{0}{0} \\
                    \bottomrule
                \end{tabular}
                }
            \end{table*}
            
            We evaluate all seven configurations (M1--M7) under six composite transport chains, using 100 files per configuration. Table~\ref{tab:exp2b-layered} reports recovery rates. For multi-layer configurations (M5--M7), we show both per-layer results and the union (any-layer) outcome. From the results, we draw the following observations.

        \begin{itemize}[noitemsep,nolistsep,leftmargin=*]
            \item \textbf{Symbolic stacking (M5) adds redundancy without interference:} M5 (WS+ZW+HG) achieves the same composite-chain coverage as its strongest component, HG. Detection remains 100\% through Tier-2 and Tier-1+2, while WS and ZW are eliminated by those chains. Per-layer results match the standalone baselines M1--M3 exactly, confirming that stacking WS, ZW, and HG on disjoint surfaces does not alter individual method behavior. At Tier-3 and beyond, all layers fail. 

            \item \textbf{Hybrid Stack (M6) extends coverage through Tier-3:} M6 (ZW+HG+LM) is the only configuration that preserves detection under adversarial non-semantic processing, achieving 97\% recovery at Tier-3 and Tier-1+2+3 via the linguistic layer. At Tier-1, all three layers recover independently (99\%); at Tier-2 and Tier-1+2, ZW fails but HG and LM remain intact (99\%). For every chain, M6-ANY matches the LM per-layer result, showing that the any-layer-recovers policy does not change the final detection outcome while still providing early-tier redundancy.
        
            \item \textbf{M7 exposes cross-layer interference:} Adding WS to the hybrid stack (M7: WS+ZW+HG+LM) causes the LM layer to collapse to 0\% at Tier-3, whereas M6 retains 97\%. This is due to T10 since WS replaces existing ASCII spaces with Unicode space variants rather than inserting new characters, and stripping those variants removes bytes that the linguistic decoder requires for token alignment. By excluding WS, M6 leaves only ZW insertions and HG substitutions above the linguistic text, and both are byte-exact-invertible after stripping (Section~\ref{sec:stacking}). This is a meaningful deviation from standalone behavior as M7's Tier-3 LM failure reflects true cross-layer interference rather than a general stacking artifact.
            
            \item \textbf{M7 exhibits fragile chained recovery:} Although M7's LM layer fails completely under isolated Tier-3, it recovers 96\% under the cumulative Tier-1+2+3 chain. This occurs because T05 (NFKC normalization) in Tier-2 converts WS Unicode spaces back to ASCII before T10 executes, preventing the byte deletions that otherwise break LM decoding. This confirms the transform-order dependency predicted in Section~\ref{sec:taxonomy} and shows that M7's recovery depends on a favorable chain ordering an adversary could avoid.

        \end{itemize}

    \subsection{Deployment Feasibility} \label{sec:results_practical}

        \subsubsection{False-Positive Rejection} \label{sec:results_exp3}

            \begin{figure}[t]
            \centering
            \newcommand{\statbox}[3]{%
                \begin{minipage}[t]{0.22\linewidth}
                    \begin{tcolorbox}[statcard]
                        \centering
                        \small\textbf{#1}\\[1pt]
                        \footnotesize\textit{#2}\\[3pt]
                        \large #3
                    \end{tcolorbox}
                \end{minipage}%
            }
            \statbox{HMAC}{0 / 400 decodes}{$P_{\mathrm{fp}} \leq 2.9{\times}10^{-32}$}%
            \hfill
            \statbox{EdDSA}{0 / 400 decodes}{$P_{\mathrm{fp}} \leq 2.9{\times}10^{-36}$}%
            \hfill
            \statbox{LM Candidates}{both schemes}{200/200 non-None}%
            \hfill
            \statbox{LM Rejection Rate}{post-verification}{200/200 rejected}
            \caption{False-positive analysis (4 decoders M1--M4 $\times$ 100 unmarked texts from prose dataset, both schemes). HMAC tested against a $10^{7}$-token registry and EdDSA tested against $10^{3}$ public keys yielding zero matches.}
            \label{fig:exp3-fp}
            \end{figure}

            We run each of the four decoders on 100 unmarked prose documents under both verification schemes: HMAC against a $10^{7}$-token registry ($10^{3}$~organization keys $\times$ $10^{4}$~file IDs), and EdDSA against $10^{3}$~registered public keys. Figure~\ref{fig:exp3-fp} reports the results.
            Zero verified matches were observed across all 800 trials. The linguistic decoder always produces output given any token sequence, returning a non-\texttt{None} candidate on all 200 documents (100 per scheme), but the extracted bytes are effectively random and fail verification under both schemes.
            
            For HMAC, a random 128-bit candidate matches any entry in a $10^{7}$-token registry with probability $P_{\mathrm{fp}} \leq 10^{7} \times 2^{-128} \approx 2.9 \times 10^{-32}$ per trial. For EdDSA, a random 68-byte payload constitutes a valid signature with probability ${\approx}\,2^{-128}$ per public key; across $10^{3}$~registered keys, $P_{\mathrm{fp}} \leq 10^{3} \times 2^{-128} \approx 2.9 \times 10^{-36}$. The EdDSA bound is four orders of magnitude tighter since the number of registered public keys ($10^{3}$) is far smaller than the HMAC registry ($10^{7}$), and the verification is per-key.

            \begin{table}[t]
            \centering
            \caption{Encoding and decoding latency per verification scheme (100 files per configuration). Symbolic methods run on CPU in sub-millisecond time for both schemes; linguistic configurations are GPU-bound and scale with payload size. The HMAC scheme (18-byte framed payload) requires fewer SAAC tokens; the EdDSA scheme (70-byte payload) requires proportionally more.}
            \label{tab:exp4-timing}
            \setlength\extrarowheight{-4pt}
                \resizebox{\linewidth}{!}{
                \begin{tabular}{@{}llccccrr@{}}
                    \toprule
                    & \textbf{Configuration}
                        & \textbf{Enc.\ mean$\pm$std} & \textbf{Enc.\ max}
                        & \textbf{Dec.\ mean$\pm$std} & \textbf{Dec.\ max}
                        & \textbf{Enc.\,OK} & \textbf{Dec.\,OK} \\
                    \midrule
                    \multicolumn{8}{@{}l}{\textit{\textbf{HMAC-SHA256} (18-byte framed payload)}} \\
                    \cmidrule{2-8}
                    \multirow{4}{*}{\rotatebox{90}{\scriptsize \texttt{CPU} \textbf{(ms)}}}
                    & M1 (WS)         & $0.149 \pm 0.017$ & 0.311 & $0.172 \pm 0.020$ & 0.366 & 100\% & 100\% \\
                    & M2 (ZW)         & $0.298 \pm 0.008$ & 0.330 & $0.146 \pm 0.005$ & 0.170 & 100\% & 100\% \\
                    & M3 (HG)         & $0.213 \pm 0.009$ & 0.258 & $0.188 \pm 0.007$ & 0.251 & 100\% & 100\% \\
                    & \textbf{M5} (WS+ZW+HG)   & $0.706 \pm 0.009$ & 0.747 & $0.536 \pm 0.009$ & 0.582 & 100\% & 100\% \\
                    \cmidrule{2-8}
                    \multirow{3}{*}{\rotatebox{90}{\scriptsize \texttt{GPU} \textbf{(ms)}}}
                    & M4 (LM)           & $1{,}151 \pm 780$ & $5{,}160$ & $172 \pm 68$ & $329$ & 100\% & 99\% \\
                    & \textbf{M6} (ZW+HG+LM)    & $1{,}403 \pm 1{,}321$ & $8{,}860$ & $194 \pm 71$ & $317$ & 96\% & 96\% \\
                    & M7 (WS+ZW+HG+LM) & $1{,}418 \pm 1{,}625$ & $9{,}901$ & $184 \pm 70$ & $374$ & 97\% & 97\% \\
                    \midrule
                    \multicolumn{8}{@{}l}{\textit{\textbf{EdDSA / Ed25519} (70-byte framed payload)}} \\
                    \cmidrule{2-8}
                    \multirow{4}{*}{\rotatebox{90}{\scriptsize \texttt{CPU} \textbf{(ms)}}}
                    & M1 (WS)         & $0.170 \pm 0.014$ & 0.294 & $0.208 \pm 0.005$ & 0.223 & 100\% & 100\% \\
                    & M2 (ZW)         & $0.350 \pm 0.011$ & 0.391 & $0.167 \pm 0.005$ & 0.179 & 100\% & 100\% \\
                    & M3 (HG)         & $0.238 \pm 0.011$ & 0.277 & $0.207 \pm 0.006$ & 0.242 & 100\% & 100\% \\
                    & \textbf{M5} (WS+ZW+HG)   & $0.898 \pm 0.016$ & 0.954 & $0.693 \pm 0.016$ & 0.820 & 100\% & 100\% \\
                    \cmidrule{2-8}
                    \multirow{3}{*}{\rotatebox{90}{\scriptsize \texttt{GPU} \textbf{(ms)}}}
                    & M4 (LM)           & $3{,}584 \pm 821$ & $7{,}490$ & $512 \pm 119$ & $1{,}348$ & 99\% & 82\% \\
                    & \textbf{M6} (ZW+HG+LM)    & $3{,}959 \pm 1{,}191$ & $9{,}703$ & $508 \pm 84$ & $784$ & 95\% & 95\% \\
                    & M7 (WS+ZW+HG+LM) & $3{,}783 \pm 966$ & $8{,}800$ & $535 \pm 121$ & $1{,}353$ & 99\% & 99\% \\
                    \bottomrule
                \end{tabular}
                }
            \end{table}

            \subsubsection{Computational Overhead} \label{sec:results_exp4}
            We measure encoding and decoding wall-clock latency on the test hardware (Section~\ref{sec:evironment}) for all seven configurations across 100 files under both verification schemes. Table~\ref{tab:exp4-timing} reports aggregate statistics.
            Symbolic methods (M1--M3, M5) are sub-millisecond on CPU under both schemes, with EdDSA payloads adding 15--30\% to encode/decode time (e.g., M5: 0.90\,ms vs.\ 0.71\,ms encoding), negligible in absolute terms. The linguistic method dominates timing and scales with payload size: under HMAC, M4 encodes in ${\sim}$1.2\,s and decodes in ${\sim}$0.17\,s; under EdDSA, encoding rises to ${\sim}$3.6\,s and decoding to ${\sim}$0.51\,s, reflecting the ${\sim}3.9\times$ larger framed payload requiring proportionally more SAAC tokens. The self-delimiting length prefix enables the decoder to terminate once the embedded identifier is fully recovered, making decode substantially faster than encode for both schemes. Cross-class configurations (M6, M7) add negligible overhead beyond M4 standalone.
            EdDSA also exhibits lower success rates for the linguistic method, with M4 decode success at 82\% compared to 99\% under HMAC. The ${\sim}3.9\times$ larger EdDSA payload (70 vs.\ 18~bytes) requires proportionally more SAAC tokens and thus more GPT-2 generated text. This longer text increases the probability that GPT-2's byte-level BPE vocabulary emits tokens containing non-ASCII Unicode characters (e.g., multi-byte sequences or rare glyphs); these tokens encode correctly during the SAAC forward pass, but when downstream symbolic layers are subsequently stripped before decoding, the byte-stream modifications cause the arithmetic-coding interval to desynchronize, producing a cascade failure. Because this is an artifact of GPT-2's 50{,}257-token vocabulary rather than a fundamental limitation of arithmetic-coding steganography, a production deployment using a larger model with a cleaner vocabulary would substantially reduce these failures (Section~\ref{sec:discussion}). For stacked configurations (M6, M7), encode failures (4--5\% under HMAC, 1--5\% under EdDSA) arise when the linguistically generated text is too short to provide sufficient capacity for the symbolic layers above it.
            From a deployment perspective, canary seeding is a one-time organizational cost as Mode~A (M5) marks 500 files in under 0.5\,s and Mode~B (M6) requires approximately 12~minutes (HMAC) to 33~minutes (EdDSA) of GPU time for the same count, acceptable as a batch operation. Vendor-side detection is the latency-sensitive path where symbolic-only decoding (M5) adds under 1\,ms per inbound request. Hybrid decoding (M6) adds ${\sim}$0.19\,s (HMAC) to ${\sim}$0.51\,s (EdDSA), dominated by the LM forward pass. An early-termination policy that stops at the first verified symbolic match reduces this to sub-millisecond latency when symbolic layers survive transport (Section~\ref{sec:discussion}).

    \subsection{End-to-End Detection in an AI-Ransomware Scenario} \label{sec:results_e2e}

\begin{figure}
            \centering
            \includegraphics[width=1\linewidth]{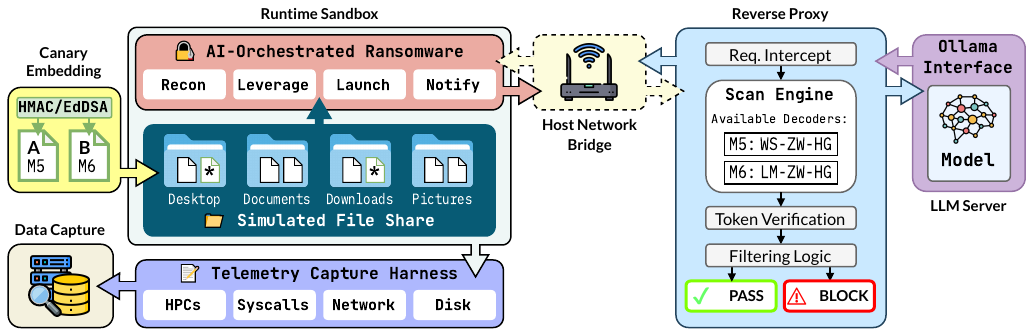}
            \caption{End-to-end ransomware case study experimental infrastructure, where the sample is run within a sandbox collecting relevant telemetry, and the reverse proxy scans inputs before they are input to the LLM.}
            \label{fig:rware_reverse_proxy}
        \end{figure}

        We validate the framework beyond the transport experiments by executing the AI-ransomware exfiltration scenario from Section~\ref{sec:threat_model} using a \textsc{Ransomware~3.0}-style attack pipeline in an instrumented sandbox, with setup and flow of data given in Figure~\ref{fig:rware_reverse_proxy}.
        %
        The experiment exercises the complete canary lifecycle from simulated organizational seeding through adversarial file targeting and vendor-side detection using the reverse proxy deployment described in Section~\ref{sec:detection}, using the HMAC verification scheme. This validates end-to-end detection under Tier-0 transport conditions where the ransomware submits raw extracted text from files. The ablations presented in Sections~\ref{sec:results_exp1}--\ref{sec:results_exp2b} establish robustness bounds under higher-tier transport transformations.
        
        \subsubsection{Scenario and Instrumentation}  \label{sec:e2e_scenario}

            We construct a simulated file share containing 288 files within a QEMU/KVM virtual machine running Ubuntu~20.04 (8\,GB RAM, 4~vCPUs). Among these, 10 pseudo-sensitive files are seeded, and 2 are selected as canary candidates (${\sim}0.7\%$ density). Mode~A canaries (M5: WS+ZW+HG) are embedded into existing sensitive-looking documents and Mode~B canaries (M6: ZW+HG+LM) are generated as standalone files with filenames designed to attract adversarial interest (e.g., \texttt{Crypto\_Wallet\_Seed\_Phrase.txt}, \texttt{Dropbox\_API\_Token.txt}).
            The attack agent is a \textsc{Ransomware~3.0}-style Go binary~\cite{raz2025ransomware30} that queries a locally hosted LLM (GPT-OSS 20B via \textsc{Ollama}) to generate Lua scripts for each lifecycle phase. All LLM API traffic passes through the reverse proxy (Section~\ref{sec:detection}), which runs the decoder suite against the canary registry before forwarding requests to the model.
            Four telemetry channels are simultaneously captured including network traffic,
            disk I/O, 
            syscall activity,
            and hardware performance counters
            at sub-second temporal resolution. We execute three run configurations: an undefended baseline ransomware run executes with the proxy in passthrough mode (no canary manifest, no scanning) and aims to establish the full attack lifecycle; an M5 defended run executes the ransomware with the proxy configured for Mode~A detection; and an M6 defended run executes the ransomware with the proxy configured for Mode~B detection.

            \begin{figure}[t]
                \centering
                \includegraphics[width=\linewidth]{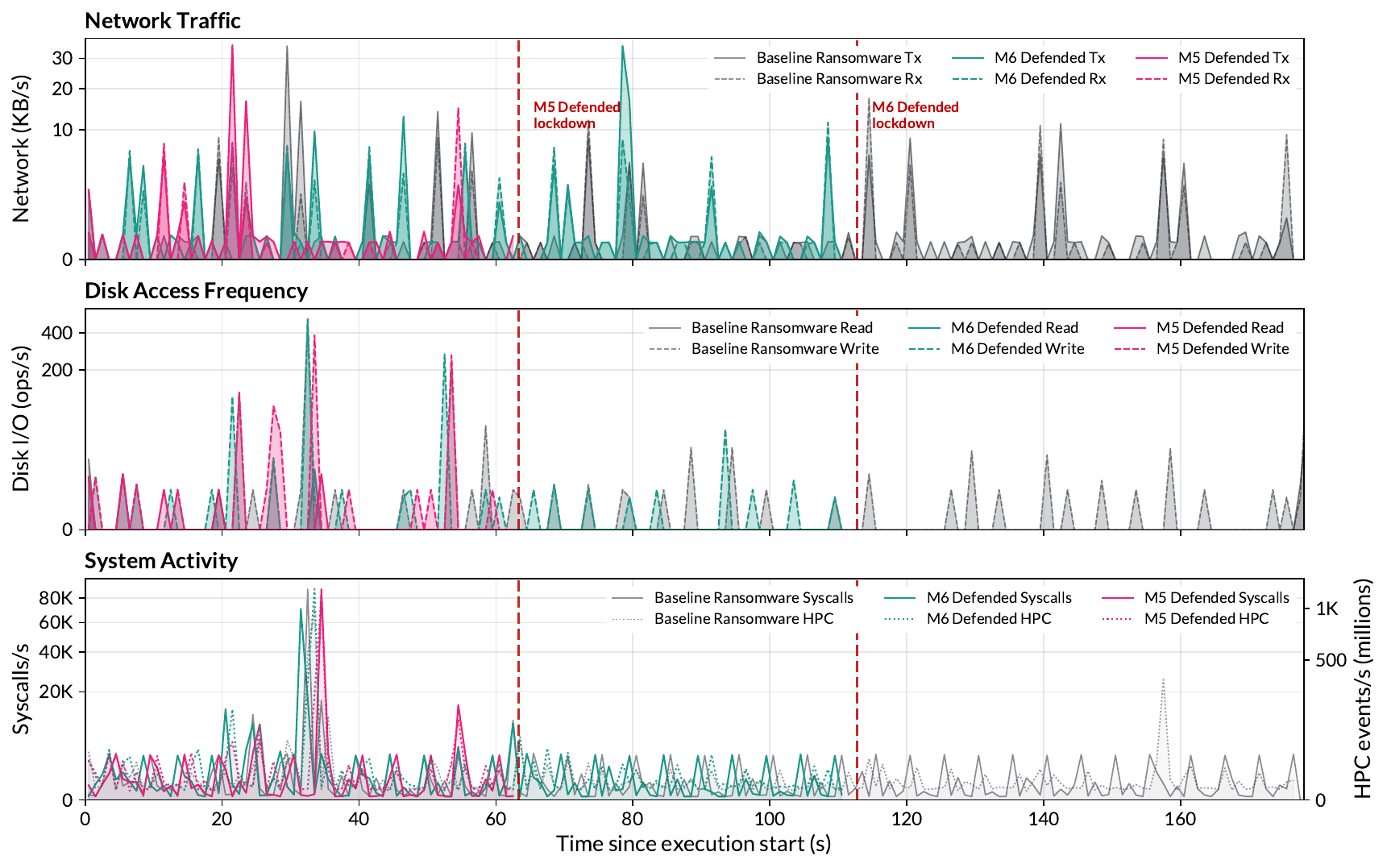}
                \caption{Multi-channel telemetry overlay for the three experiment runs. Network traffic (top), disk I/O read/write frequency (middle), and system activity including syscall rate and hardware performance counters (bottom) are shown with a shared time axis. Red dashed vertical lines mark the proxy lockdown events for the M5- and M6-defended runs. The undefended ransomware trace extends across the full execution window, while the defended traces terminate shortly after detection.}
                \label{fig:e2e-telemetry}
            \end{figure}

        \subsubsection{Attack Lifecycle Characterization} \label{sec:e2e_lifecycle}
    
            The undefended run provides the behavioral ground truth for the \textsc{Ransomware~3.0} attack lifecycle, which proceeds through seven distinct phases identifiable in the program log and corroborated by the telemetry channels (Figure~\ref{fig:e2e-telemetry}):
        
            \begin{enumerate}[noitemsep,nolistsep,leftmargin=*]
                \item \textbf{Probe:} performs recon via Lua script to identify OS, username, hostname, and working directory.
                \item \textbf{Scan:} writes code to enumerate the entirety of the file share, discovering all 288 files.
                \item \textbf{Target:} selects 10 high-value files from the file share based on filename heuristics.
                \item \textbf{Extract:} reads target file contents and submits them to the LLM for leverage assessment.
                \item \textbf{Decide:} decides that the files have leverage value and selects encryption as the payload.
                \item \textbf{Encrypt:} generates and executes the encryption payload on targeted files.
                \item \textbf{Note:} generates the ransom note, completing the full attack lifecycle.
            \end{enumerate}
        
            The agent employs a phase retry loop (up to 10 iterations) in which the LLM generates a Lua script, executes it, then validates the output, retrying with corrective feedback on failure. Because LLM code generation is non-deterministic, some phases require multiple iterations. Per-phase iteration counts, elapsed times, and total LLM calls for all three runs are compared in Table~\ref{tab:e2e-phases}.

            \begin{table}[t]
                \centering\small
                \caption{Per-phase iteration counts and elapsed time for the three ransomware runs. Each iteration involves one or more LLM calls to generate and/or validate a Lua script; elapsed times denote wall-clock duration per phase. $^{*}$Additional iterations due to non-deterministic LLM code-generation or validation failures. \colorbox{blockedBg}{\textcolor{blockedFg}{\textsc{blocked}}}~= proxy lockdown prevented phase execution; \colorbox{partialBg}{\textsc{blocked}}~= phase partially executed (request sent) but halted.}
                \label{tab:e2e-phases}
                \resizebox{\linewidth}{!}{%
                \begin{tabular}{@{}ll rrrrrrr c@{}}
                    \toprule
                    & & \textbf{Probe} & \textbf{Scan} & \textbf{Target} & \textbf{Extract} & \textbf{Decide} & \textbf{Encrypt} & \textbf{Note} & \textbf{Total} \\
                    \midrule
                    \multirow{3}{*}{\rotatebox[origin=c]{90}{\scriptsize\textbf{Iterations}}}
                        & Baseline       & 1        & 1        & 1       & 2$^{*}$      & 1          & 3$^{*}$    & 1          & 17 calls / 7 scripts \\
                        & Ransomware + M5 & 1        & 1        & 1       & \PARTIAL{1}  & \BLOCKED   & ---        & \BLOCKED   & 6 calls / 3 scripts  \\
                        & Ransomware + M6 & 5$^{*}$  & 2$^{*}$  & 1       & \PARTIAL{1}  & \BLOCKED   & ---        & \BLOCKED   & 16 calls / 8 scripts \\
                    \midrule
                    \multirow{3}{*}{\rotatebox[origin=c]{90}{\scriptsize\textbf{Time (s)}}}
                        & Baseline        & 23        &8         & 10         & 39       & 2         & 78         & 15         & ${\sim}$177 \\
                        & Ransomware + M5 & 15         & 8        & 32         & 9        & \BLOCKED  & ---        & \BLOCKED   & ${\sim}$64    \\
                        & Ransomware + M6 & 60         & 20       & 12         & 21       & \BLOCKED  & ---        & \BLOCKED   & ${\sim}$113          \\
                    \bottomrule
                \end{tabular}
                }%
            \end{table}

            \begin{table}[h]
                \centering\small
                \caption{End-to-end ransomware case study: summary of experiment outcomes.
                $\dagger$~Proxy entered lockdown after detection; all subsequent LLM calls
                and attack phases were rejected.}
                \label{tab:e2e-summary}
                \resizebox{\linewidth}{!}{%
                \begin{tabular}{@{}lrrrrccc@{}}
                    \toprule
                    \textbf{Run}
                        & \textbf{Dur.\ (s)}
                        & \textbf{LLM}
                        & \textbf{Scripts}
                        & \textbf{Traffic}
                        & \textbf{Lockdown}
                        & \textbf{Method}
                        & \textbf{Scan} \\
                        &
                        & \textbf{Calls}
                        & \textbf{Gen.}
                        & \textbf{(KB)}
                        & \textbf{Time (s)}
                        & \textbf{Chain}
                        & \textbf{Time} \\
                    \midrule
                    \rowcolor{rowBaseline}
                    Ransomware Baseline
                        & 177 & 17 & 7 & 269.0 & --- & --- & --- \\
                    \rowcolor{rowDefended}
                    Ransomware + M5$^{\dagger}$
                        & 66 & 6 & 3 & 124.7 & 63.3 & WS\,$\to$\,ZW\,$\to$\,HG\, & 0.2\,ms\,$\to$\,0.4\,ms\,$\to$\,0.7\,ms \\
                    \rowcolor{rowDefended}
                    Ransomware + M6$^{\dagger}$
                        & 116 & 16 & 8 & 237.0 & 112.8 & ZW\,$\to$\,HG\,$\to$\,LM & 0.2\,ms\,$\to$\,0.4\,ms\,$\to$\,168.3\,ms \\
                    \midrule
                    \multicolumn{8}{@{}l}{%
                        \footnotesize\textit{$\Delta$ M5 vs.\ baseline:
                        $-$111\,s ($-63\%$), $-$11 LLM calls, $-$144.3\,KB traffic}} \qquad \footnotesize\textit{$\Delta$ M6 vs.\ baseline:
                        $-$61\,s ($-34\%$), $-$1 LLM call, $-$32.0\,KB traffic} \\
                    \bottomrule
                \end{tabular}    
                }%
            \end{table}
        
        \subsubsection{Defense Intervention} \label{sec:e2e_defense}

            In both defended runs, the defense is invisible until the Extract phase submits canary-bearing content to the LLM.  The variation in pre-detection LLM calls between M5 (6 calls) and M6 (16 calls) is attributed to non-deterministic code-generation retries in the early phases, and we do not optimize the ransomware to execute in the least tries. In both defended runs, the defense intervened at the first Extract request that included canary-bearing content. Tables~\ref{tab:e2e-phases} and~\ref{tab:e2e-summary} summarize the phase-level and quantitative outcomes across all three runs.

            \begin{itemize}[style=unboxed,leftmargin=*,nosep]

                \item \textbf{Mode~A (M5: WS+ZW+HG):}
                The proxy detected the M5 canary at $t=63.3$\,s on the first extraction request containing encoded plaintext in the chat request. The proxy decoded the full symbolic chain to verify all three layers, where WS recovered the HMAC token in 0.2\,ms, ZW verified in 0.4\,ms, and HG verified in 0.7\,ms. In a production deployment, early-termination policy would stop at the WS match, but running the full chain here confirms that all three encoding surfaces survived the ransomware's prompt wrapping intact. Once in lockdown, the proxy rejected all subsequent requests (2 blocked), where the Decide and Note phases each received empty LLM responses and failed, preventing any encryption or ransom-note generation and ultimately ending the ransomware process.
    
                \item \textbf{Mode~B (M6: ZW+HG+LM):}
                The proxy detected the M6 canary at $t=112.8$\,s via the full decoder chain (ZW$\to$HG$\to$LM) operating on the LLM request payload. The \textsc{Ransomware~3.0} agent wraps extracted file content in instructional prompts (e.g., ``\textit{Analyze the following content for potential leverage\ldots}''), and the proxy's region-aware extraction isolated the document text within known delimiters before running the decoder chain. The symbolic decoders resolved quickly (ZW in 0.2\,ms, HG in 0.4\,ms) and the linguistic decoder confirmed the canary in 168.3\,ms via the SAAC decoding pass. This is consistent with the HMAC decode latencies measured in Section~\ref{sec:results_exp4}. As with M5, all three encoding layers independently confirm the embedded token, and lockdown cascade prevented the attack from progressing to encryption.
    
            \end{itemize}

            \subsubsection{Case Study Observations} \label{sec:e2e_observations}

                \begin{enumerate}[noitemsep,nolistsep,leftmargin=*]
                    \item \textbf{Early ransomware detection:} Both Mode A and B were detected during the file-analysis stage of the \textsc{Ransomware~3.0} lifecycle before progression to encryption or extortion. In the threat model of Section~\ref{sec:threat_model}, this corresponds to detection at the LLM ingestion boundary, validating the claim that AI-assisted adversaries create a viable detection interception point by their dependence on external model access. The proxy lockdown terminated the attack at the earliest possible point in the attack chain, with no files encrypted in either defended run.
                
                    \item \textbf{Sub-millisecond lockdown via early termination:} We configured the proxy to execute the complete decoder chain in both runs rather than stopping at the first positive match, in order to verify that all encoding layers survived the adversarial pipeline. For M5, all three symbolic decoders confirmed the canary in under 1\,ms total (WS\,0.2\,ms, ZW\,0.4\,ms, HG\,0.7\,ms). For M6, the two symbolic decoders resolved in under 1\,ms before the linguistic decoder confirmed the payload in 168.3\,ms. Since a single symbolic match already provides negligible false-positive risk (Section~\ref{sec:results_exp3}), a production deployment can safely stop at the first confirmed extraction. This early-termination policy would reduce both M5 and M6 detection to sub-millisecond latency, confirming the practical optimization noted in Section~\ref{sec:results_exp4}.
                
                    \item \textbf{Prompt wrapping preserves symbolic encodings:} The ransomware agent wraps extracted file content in instructional prompts before submission, which is standard practice often employed to prevent prompt injection. This wrapping prepends and appends tokens but does not modify the document body, leaving all symbolic encoding surfaces intact. For M6 canaries, the proxy's region-aware extraction further isolates the document content from surrounding prompt text via delimiter scanning, ensuring the ZW and HG decoders operated on clean input.

                \end{enumerate}                   
\section{Discussion} \label{sec:discussion}

    \subsection{Summary of Findings} \label{sec:key-findings}

        \subsubsection{Feasibility and Binary Recovery}        
            All four embedding methods achieve 100\% recovery at Tier-0 across all tested documents, confirming that a per-file cryptographic identifier can be reliably embedded and recovered from realistic plaintext (\textbf{RQ1}). A structural property which emerges from the ablation is that canary token recovery is strictly binary across all methods and transforms. That is, a transform either preserves the encoding surface entirely or destroys it completely, with no correctable partial corruption surface per file. This follows from the global nature of the transforms (operating uniformly on character classes) and the cascading structure of arithmetic coding errors in the linguistic method. Error-correction coding therefore provides no benefit, as there are no near-miss cases for redundancy to bridge. A naive approach of repeating the token within the same encoding surface may aid in increasing chances of recovery in chunked copy-pastes, but does not help with survival as all copies are lost simultaneously per encoding surface. Due to this, method diversity (i.e., hybrid stacking) rather than redundancy increases robustness (\textbf{RQ2}).

        \subsubsection{Orthogonal Failure Surfaces}
            
            The per-transform ablation (Section~\ref{sec:results_exp1}) reveals orthogonal failure surfaces since WS fails at Tier~2 (whitespace normalization), ZW at Tier~2--3 (format-character stripping), HG at Tier~3 (confusable normalization), and LM only at Tier~4 (semantic rewriting). Within the symbolic class, T11 (punctuation/case stripping) exposes a finer distinction where WS and ZW survive because their decoders search for specific codepoints regardless of surrounding content and HG fails because case-folding removes uppercase-only confusable pairs from the eligible set. A consequential structural finding is the inverse coverage profile between T10 and T11. T10 destroys all symbolic methods but LM survives at 98\%, while T11 destroys LM and nearly destroys HG (8\%) but WS and ZW survive at 100\%. This bidirectional complementarity between method classes is the empirical foundation of defense-in-depth through stacking (\textbf{RQ2}).

        \subsubsection{Principled Composition for Defense-in-Depth}
        
            The stacking experiments (Section~\ref{sec:results_exp2b}) confirm that method diversity provides broader coverage, but only under principled composition (\textbf{RQ3}). M5 (WS+ZW+HG) provides triple redundancy at Tier~1 and extends coverage through Tier~2 at 100\% via HG. M6 (ZW+HG+LM) further extends detection through Tier~3 at 97\% via the linguistic layer, which is the only configuration maintaining detection under adversarial non-semantic processing. The M7 versus M6 comparison shows that including WS in a cross-class stack reduces Tier~3 coverage from 97\% to 0\% due to T10 deleting WS replacement artifacts and corrupting the linguistic decoder's byte stream. This produces a design principle where insertion-based (ZW) or bijective-substitution (HG) symbolic methods should be layered on linguistic cover text. The two deployment modes are complementary since Mode~A (M5) covers T11 via WS and ZW and Mode~B (M6) covers Tier~3 via the linguistic layer. Only Tier~4 defeats both modes simultaneously (\textbf{RQ3}).

        \subsubsection{End-to-End Ransomware Detection}
    
            This case study (Section~\ref{sec:results_e2e}) validates the canary lifecycle under a realistic AI-assisted attack workflow (\textbf{RQ5}). Both deployment modes were detected during the Extract phase of the \textsc{Ransomware~3.0} lifecycle and the proxy lockdown prevented file encryption. With the self-delimiting payload framing, the linguistic decoder completes in 168.3\,ms for the HMAC scheme by terminating once the identifier is fully recovered. Early termination at the first symbolic match would reduce detection to sub-millisecond latency in both cases (\textbf{RQ4}). The controlled stacking experiments (Section~\ref{sec:results_exp2b}) establish that these capabilities extend to higher adversarial tiers since M5 provides coverage through Tier-2 and M6 through Tier-3.

    \subsection{Security Implications} \label{sec:security-implications}
        
        \begin{itemize}[noitemsep,nolistsep,leftmargin=*]
        
            \item \textbf{Detection at the processing boundary:}
            AI-powered malware including agentic ransomware, espionage agents, credential harvesters, and social-engineering tools cannot execute intelligent attacks without first submitting stolen files to an external LLM for analysis. Our results demonstrate that this shared dependency creates a viable detection opportunity at the processing boundary where the vendor-side filter identifies canary documents before model processing, with zero false positives under both shared-key verification and public-key verification. From the vendor viewpoint, the canary identifier reveals nothing about file content and thus preserves privacy. The end-to-end case study further validates detection against LLM-orchestrated ransomware, the mechanism operates at the interception point common to all four AI-malware categories, as the detection is content-based rather than attack-specific. In the case of agentic ransomware, detection occurs during the reconnaissance phase (before encryption, exfiltration, or ransom-note generation) and provides early opportunities for mitigation.
        
            \item \textbf{Adversarial cost asymmetry:}
            An attacker suspecting canaries must apply aggressive normalization to all exfiltrated files since canaries are visually indistinguishable from legitimate documents. Each sanitization step degrades utility: NFKC normalization alters code semantics, full ASCII stripping destroys non-English text and formatting, and LLM paraphrasing changes wording and introduces stylistic artifacts (see Section~\ref{sec:local-llm}). The defender forces a choice between accepting detection risk and degrading the stolen data.
        
            \item \textbf{Insider threat deterrence:}
            For the secondary insider-upload scenario, the canaries change the risk calculus for authorized users, creating a deterrence effect analogous to traditional honeytokens but with detection at the AI service boundary rather than through callback infrastructure.

        \end{itemize}

    \subsection{Practical Considerations} \label{sec:practical-considerations}

        \subsubsection{Encode \& Decode Overhead}
        Symbolic seeding (Mode~A) is effectively free at organizational scale as M5 encodes at under 1\,ms per file, completing 10{,}000 files in under 10~seconds on a consumer-grade CPU. Linguistic generation (Mode~B) requires approximately 1.4\,s (HMAC) to 4.0\,s (EdDSA) per file on GPU, or roughly a one-time batch cost of 4--11~hours for 10{,}000 generated canaries. At a practical seeding density of 1--5\%, an organization with 100{,}000 files generates 1{,}000--5{,}000 canaries, well within these budgets.
        In terms of decoding, Symbolic-only adds under 1\,ms per request while hybrid decoding (M6) adds ${\sim}$0.19\,s (HMAC) to ${\sim}$0.51\,s (EdDSA). Because symbolic layers decode first and the any-layer-recovers policy triggers on the first verified extraction, an early-termination policy can skip the linguistic decoder when symbolic extraction already yields a confirmed match, reducing average per-request latency to sub-millisecond levels while preserving full-stack coverage for Mode~B canaries under higher-tier transport.

        \subsubsection{Key Maintenance \& Verification Schemes}
        For HMAC, the vendor maintains a pre-loaded token registry which is updated (via secure transmission, API, or desktop application) as canaries are added or rotated (e.g. every two weeks). For EdDSA, the vendor stores only the organization's 32-byte public key with no shared secrets and no per-file registry updates, allowing any canary to be verifiable on-the-fly. This eliminates the key-management and registry-synchronization burden at the cost of a larger embedded payload (70~vs.\ 18~bytes framed), increasing symbolic capacity requirements and linguistic encoding time.

    \subsection{Limitations} \label{sec:limitations}

        \subsubsection{Semantic Rewriting and Local Adversarial LLMs} \label{sec:local-llm}
        
            Tier~4 semantic rewriting defeats all embedding methods simultaneously and is identified as the hard limit of the current framework. Robustness against paraphrasing is not achievable with existing steganographic techniques, since rewriting intentionally destroys both character-level and token-level signals. We argue this is operationally costly for adversaries, but do not claim resilience against it. The T11 vulnerability is related but distinct since punctuation/case stripping defeats LM due to BPE tokenization and HG due to uppercase-only pair removal, while WS and ZW survive. A case-invariant linguistic encoder or reduced HG mapping restricted to case-stable pairs could potentially close this gap.
            
            An adversary rewriting every exfiltrated file via a local LLM before cloud upload pays three costs: (i)~per-file inference latency during time-sensitive operations (even a small model at ~100ms/file across thousands of files adds meaningful delay to ransomware's file-triage phase), (ii)~content degradation (small models produce lossy rewrites that can destroy the exact information the adversary needs, such as credential strings, configuration values, code semantics), and (iii)~expanded detection surface (model weights on disk, GPU utilization spikes, and suspicious process trees are exactly the behavioral signals endpoint tools are designed to catch). The defense thus forces the adversary into a costlier, more detectable posture even when canary files are suspected.

        \subsubsection{Linguistic Method Constraints}
        
            SAAC is sequential and each token requires a full GPT-2 forward pass, and so the arithmetic coding interval at step~$t$ depends on the token selected at step~$t{-}1$, causing encoding latency to scales with payload size. This latency is irreducible without model distillation or architectural changes. GPT-2's byte-level BPE vocabulary occasionally produces tokens containing non-ASCII Unicode characters; these encode correctly during the SAAC forward pass but can cause arithmetic-coding desynchronization during decoding when the intervening symbolic layers are stripped. Because the EdDSA payload requires ${\sim}3.9\times$ more generated text than HMAC, the probability of encountering such a token rises, resulting in 82\% decode success for M4 under EdDSA vs.\ 99\% under HMAC. This is a GPT-2 vocabulary artifact, not a fundamental limitation of the arithmetic-coding mechanism. The SAAC scheme is model-agnostic and the framework supports model substitution via its module interface (Section~\ref{sec:funcs}), and a production deployment using a larger model with a BPE vocabulary without multi-byte non-ASCII tokens) would substantially reduce these failures and also improve embedding rates.
        
        \subsubsection{Scope and Evaluation Boundaries}
        
            Several boundaries constrain the generalizability of our results. 
            Our (Tier 1--4) transforms are deterministic approximations and can vary in real-world transport and ingestion workflows. 
            The cross-class composition constraint (WS excluded from Mode~B) reduces Mode~B from four channels to three, though the practical impact is minimal since WS provides no coverage beyond ZW, and the linguistic layer provides the Tier~3 coverage no symbolic method can. Finally, the framework requires a cooperating vendor or enterprise proxy and the framework provides no detection capability in its absence.

\section{Conclusion} \label{sec:conclusion}

    The growing reliance of both enterprise users and adversaries on cloud-hosted LLMs for document analysis has created an exfiltration channel that traditional data-loss controls do not cover. We presented a steganographic canary framework that embeds cryptographically verifiable identifiers into plaintext documents, enabling detection at the AI service ingestion boundary before model processing, under both shared-key and public-key (registry-less) verification.
    Mode~A of the framework marks existing documents with layered symbolic encodings and provides reliable detection through Tier~2 sanitization at sub-millisecond cost whereas Mode~B adds a linguistic layer to extend robustness through Tier~3 with 97\% recovery. Recovery is per-file binary, and so each transport transform either preserves or destroys an encoding surface entirely. Our evaluation further shows that improper layer composition can eliminate robustness through cross-layer interference, motivating the invertibility requirement and safe-combination principles we establish. An end-to-end case study against a \textsc{PromptLock}-style ransomware pipeline confirms that both modes detect and block canary-bearing uploads during reconnaissance, before file encryption occurs.
    Open directions include extending the framework to multi-modal canaries for non-text documents, integration with enterprise AI governance workflows, adversarial robustness analysis under active canary identification, and encoding schemes that introduce partial-corruption regimes where error correction coding becomes viable.
    To our knowledge, this is the first framework to systematically combine symbolic and linguistic text steganography into layered canary documents, establish method-agnostic composition principles, and evaluate them against a transport-threat taxonomy tailored to the LLM-upload threat model.

\bibliographystyle{ACM-Reference-Format}
\bibliography{refs}


\begin{thebibliography}{44}


\ifx \showCODEN    \undefined \def \showCODEN     #1{\unskip}     \fi
\ifx \showISBNx    \undefined \def \showISBNx     #1{\unskip}     \fi
\ifx \showISBNxiii \undefined \def \showISBNxiii  #1{\unskip}     \fi
\ifx \showISSN     \undefined \def \showISSN      #1{\unskip}     \fi
\ifx \showLCCN     \undefined \def \showLCCN      #1{\unskip}     \fi
\ifx \shownote     \undefined \def \shownote      #1{#1}          \fi
\ifx \showarticletitle \undefined \def \showarticletitle #1{#1}   \fi
\ifx \showURL      \undefined \def \showURL       {\relax}        \fi
\providecommand\bibfield[2]{#2}
\providecommand\bibinfo[2]{#2}
\providecommand\natexlab[1]{#1}
\providecommand\showeprint[2][]{arXiv:#2}

\bibitem[{330k}(2015)]%
        {330k-steg}
\bibfield{author}{\bibinfo{person}{{330k}}.} \bibinfo{year}{2015}\natexlab{}.
\newblock \bibinfo{title}{Unicode Steganography with Zero-Width Characters}.
\newblock \bibinfo{howpublished}{Online tool and {JavaScript} library}.
\newblock
\urldef\tempurl%
\url{https://330k.github.io/misc_tools/unicode_steganography.html}
\showURL{%
\tempurl}


\bibitem[Alneyadi et~al\mbox{.}(2016)]%
        {alneyadi2016dlp}
\bibfield{author}{\bibinfo{person}{Sultan Alneyadi}, \bibinfo{person}{Elankayer Sithirasenan}, {and} \bibinfo{person}{Vallipuram Muthukkumarasamy}.} \bibinfo{year}{2016}\natexlab{}.
\newblock \showarticletitle{A survey on data leakage prevention systems}.
\newblock \bibinfo{journal}{\emph{J. Netw. Comput. Appl.}} \bibinfo{volume}{62}, \bibinfo{number}{C} (\bibinfo{date}{Feb.} \bibinfo{year}{2016}), \bibinfo{pages}{137–152}.
\newblock
\showISSN{1084-8045}
\href{https://doi.org/10.1016/j.jnca.2016.01.008}{doi:\nolinkurl{10.1016/j.jnca.2016.01.008}}


\bibitem[Anand et~al\mbox{.}(2025)]%
        {anand2025larm}
\bibfield{author}{\bibinfo{person}{P~Mohan Anand}, \bibinfo{person}{PV~Sai Charan}, \bibinfo{person}{Hrushikesh Chunduri}, {and} \bibinfo{person}{Sandeep~Kumar Shukla}.} \bibinfo{year}{2025}\natexlab{}.
\newblock \showarticletitle{LARM: Linux Anti Ransomware Monitor}.
\newblock \bibinfo{journal}{\emph{Computers \& Security}} (\bibinfo{year}{2025}), \bibinfo{pages}{104700}.
\newblock


\bibitem[Bernstein(2006)]%
        {bernstein2006curve}
\bibfield{author}{\bibinfo{person}{D.J. Bernstein}.} \bibinfo{year}{2006}\natexlab{}.
\newblock \showarticletitle{Curve25519: new Diffie-Hellman speed records}. In \bibinfo{booktitle}{\emph{Public Key Cryptography - PKC 2006 (9th International Conference on Practice and Theory in Public-Key Cryptography, New York NY, USA, April 24-26, 2006, Proceedings)}} \emph{(\bibinfo{series}{Lecture Notes in Computer Science})}, \bibfield{editor}{\bibinfo{person}{M.~Yung}, \bibinfo{person}{Y.~Dodis}, \bibinfo{person}{A.~Kiayias}, {and} \bibinfo{person}{T.~Malkin}} (Eds.). \bibinfo{publisher}{Springer}, \bibinfo{address}{Germany}, \bibinfo{pages}{207--228}.
\newblock
\showISBNx{3-540-33851-9}
\href{https://doi.org/10.1007/11745853\_14}{doi:\nolinkurl{10.1007/11745853\_14}}


\bibitem[Boucher et~al\mbox{.}(2022)]%
        {boucher2022badcharacters}
\bibfield{author}{\bibinfo{person}{Nicholas Boucher}, \bibinfo{person}{Ilia Shumailov}, \bibinfo{person}{Ross Anderson}, {and} \bibinfo{person}{Nicolas Papernot}.} \bibinfo{year}{2022}\natexlab{}.
\newblock \showarticletitle{Bad Characters: Imperceptible NLP Attacks}. In \bibinfo{booktitle}{\emph{2022 IEEE Symposium on Security and Privacy (SP)}}. \bibinfo{pages}{1987--2004}.
\newblock
\href{https://doi.org/10.1109/SP46214.2022.9833641}{doi:\nolinkurl{10.1109/SP46214.2022.9833641}}


\bibitem[Bowen et~al\mbox{.}(2009)]%
        {bowen2009baiting}
\bibfield{author}{\bibinfo{person}{Brian~M. Bowen}, \bibinfo{person}{Shlomo Hershkop}, \bibinfo{person}{Angelos~D. Keromytis}, {and} \bibinfo{person}{Salvatore~J. Stolfo}.} \bibinfo{year}{2009}\natexlab{}.
\newblock \showarticletitle{Baiting Inside Attackers Using Decoy Documents}. In \bibinfo{booktitle}{\emph{Security and Privacy in Communication Networks}}, \bibfield{editor}{\bibinfo{person}{Yan Chen}, \bibinfo{person}{Tassos~D. Dimitriou}, {and} \bibinfo{person}{Jianying Zhou}} (Eds.). \bibinfo{publisher}{Springer Berlin Heidelberg}, \bibinfo{address}{Berlin, Heidelberg}, \bibinfo{pages}{51--70}.
\newblock
\showISBNx{978-3-642-05284-2}


\bibitem[Casino et~al\mbox{.}(2022)]%
        {casino2022forensics}
\bibfield{author}{\bibinfo{person}{Fran Casino}, \bibinfo{person}{Thomas~K. Dasaklis}, \bibinfo{person}{Georgios~P. Spathoulas}, \bibinfo{person}{Marios Anagnostopoulos}, \bibinfo{person}{Amrita Ghosal}, \bibinfo{person}{István Bo\"{r}o\"{c}z}, \bibinfo{person}{Agusti Solanas}, \bibinfo{person}{Mauro Conti}, {and} \bibinfo{person}{Constantinos Patsakis}.} \bibinfo{year}{2022}\natexlab{}.
\newblock \showarticletitle{Research Trends, Challenges, and Emerging Topics in Digital Forensics: A Review of Reviews}.
\newblock \bibinfo{journal}{\emph{IEEE Access}}  \bibinfo{volume}{10} (\bibinfo{year}{2022}), \bibinfo{pages}{25464--25493}.
\newblock
\href{https://doi.org/10.1109/ACCESS.2022.3154059}{doi:\nolinkurl{10.1109/ACCESS.2022.3154059}}


\bibitem[Castiglione et~al\mbox{.}(2011)]%
        {castiglione2011ooxml}
\bibfield{author}{\bibinfo{person}{Aniello Castiglione}, \bibinfo{person}{Bonaventura D'Alessio}, \bibinfo{person}{Alfredo De~Santis}, {and} \bibinfo{person}{Francesco Palmieri}.} \bibinfo{year}{2011}\natexlab{}.
\newblock \showarticletitle{New Steganographic Techniques for the OOXML File Format}. In \bibinfo{booktitle}{\emph{Availability, Reliability and Security for Business, Enterprise and Health Information Systems}}, \bibfield{editor}{\bibinfo{person}{A.~Min Tjoa}, \bibinfo{person}{Gerald Quirchmayr}, \bibinfo{person}{Ilsun You}, {and} \bibinfo{person}{Lida Xu}} (Eds.). \bibinfo{publisher}{Springer Berlin Heidelberg}, \bibinfo{address}{Berlin, Heidelberg}, \bibinfo{pages}{344--358}.
\newblock
\showISBNx{978-3-642-23300-5}


\bibitem[Cen et~al\mbox{.}(2024)]%
        {cen2024ransomware}
\bibfield{author}{\bibinfo{person}{Mingcan Cen}, \bibinfo{person}{Frank Jiang}, \bibinfo{person}{Xingsheng Qin}, \bibinfo{person}{Qinghong Jiang}, {and} \bibinfo{person}{Robin Doss}.} \bibinfo{year}{2024}\natexlab{}.
\newblock \showarticletitle{Ransomware early detection: A survey}.
\newblock \bibinfo{journal}{\emph{Comput. Netw.}} \bibinfo{volume}{239}, \bibinfo{number}{C} (\bibinfo{date}{Feb.} \bibinfo{year}{2024}), \bibinfo{numpages}{20}~pages.
\newblock
\showISSN{1389-1286}
\href{https://doi.org/10.1016/j.comnet.2023.110138}{doi:\nolinkurl{10.1016/j.comnet.2023.110138}}


\bibitem[CV et~al\mbox{.}(2020)]%
        {kurolabs2020stegcloak}
\bibfield{author}{\bibinfo{person}{Jyothishmathi CV}, \bibinfo{person}{Kandavel A}, {and} \bibinfo{person}{Mohanasundar M}.} \bibinfo{year}{2020}\natexlab{}.
\newblock \bibinfo{booktitle}{\emph{{StegCloak: Hide Secrets with Invisible Characters in Plain Text Securely Using Passwords}}}.
\newblock
\urldef\tempurl%
\url{https://github.com/KuroLabs/stegcloak}
\showURL{%
\tempurl}


\bibitem[{Cyberhaven Labs}(2025)]%
        {cyberhaven2025adoption}
\bibfield{author}{\bibinfo{person}{{Cyberhaven Labs}}.} \bibinfo{year}{2025}\natexlab{}.
\newblock \bibinfo{title}{{2025 AI Adoption and Risk Report}}.
\newblock
\urldef\tempurl%
\url{https://www.cyberhaven.com/resources/report/2025-ai-adoption-risk-report}
\showURL{%
\tempurl}


\bibitem[Dai and Cai(2019)]%
        {dai2019towards}
\bibfield{author}{\bibinfo{person}{Falcon Dai} {and} \bibinfo{person}{Zheng Cai}.} \bibinfo{year}{2019}\natexlab{}.
\newblock \showarticletitle{Towards Near-imperceptible Steganographic Text}. In \bibinfo{booktitle}{\emph{Proceedings of the 57th Annual Meeting of the Association for Computational Linguistics}}, \bibfield{editor}{\bibinfo{person}{Anna Korhonen}, \bibinfo{person}{David Traum}, {and} \bibinfo{person}{Llu{\'i}s M{\`a}rquez}} (Eds.). \bibinfo{publisher}{Association for Computational Linguistics}, \bibinfo{address}{Florence, Italy}, \bibinfo{pages}{4303--4308}.
\newblock
\href{https://doi.org/10.18653/v1/P19-1422}{doi:\nolinkurl{10.18653/v1/P19-1422}}


\bibitem[Dathathri et~al\mbox{.}(2024)]%
        {dathathri2024synthid}
\bibfield{author}{\bibinfo{person}{Sumanth Dathathri}, \bibinfo{person}{Abigail See}, \bibinfo{person}{Sumedh Ghaisas}, \bibinfo{person}{Po-Sen Huang}, \bibinfo{person}{Rob McAdam}, \bibinfo{person}{Johannes Welbl}, \bibinfo{person}{Vandana Bachani}, \bibinfo{person}{Alex Kaskasoli}, \bibinfo{person}{Robert Stanforth}, \bibinfo{person}{Tatiana Matejovicova}, \bibinfo{person}{Jamie Hayes}, \bibinfo{person}{Nidhi Vyas}, \bibinfo{person}{Majd~Al Merey}, \bibinfo{person}{Jonah Brown-Cohen}, \bibinfo{person}{Rudy Bunel}, \bibinfo{person}{Borja Balle}, \bibinfo{person}{Taylan Cemgil}, \bibinfo{person}{Zahra Ahmed}, \bibinfo{person}{Kitty Stacpoole}, \bibinfo{person}{Ilia Shumailov}, \bibinfo{person}{Ciprian Baetu}, \bibinfo{person}{Sven Gowal}, \bibinfo{person}{Demis Hassabis}, {and} \bibinfo{person}{Pushmeet Kohli}.} \bibinfo{year}{2024}\natexlab{}.
\newblock \showarticletitle{{Scalable Watermarking for Identifying Large Language Model Outputs}}.
\newblock \bibinfo{journal}{\emph{Nature}} \bibinfo{volume}{634}, \bibinfo{number}{8035} (\bibinfo{year}{2024}), \bibinfo{pages}{818--823}.
\newblock
\href{https://doi.org/10.1038/s41586-024-08025-4}{doi:\nolinkurl{10.1038/s41586-024-08025-4}}


\bibitem[Davis and Suignard(2025)]%
        {unicode2025uts39}
\bibfield{author}{\bibinfo{person}{Mark Davis} {and} \bibinfo{person}{Michel Suignard}.} \bibinfo{year}{2025}\natexlab{}.
\newblock \bibinfo{booktitle}{\emph{{Unicode Technical Standard \#39: Unicode Security Mechanisms}}}.
\newblock \bibinfo{type}{Unicode Technical Standard}~39. \bibinfo{institution}{Unicode Consortium}.
\newblock
\urldef\tempurl%
\url{https://unicode.org/reports/tr39/}
\showURL{%
\tempurl}


\bibitem[Ding et~al\mbox{.}(2023)]%
        {ding2023discop}
\bibfield{author}{\bibinfo{person}{Jinyang Ding}, \bibinfo{person}{Kejiang Chen}, \bibinfo{person}{Yaofei Wang}, \bibinfo{person}{Na Zhao}, \bibinfo{person}{Weiming Zhang}, {and} \bibinfo{person}{Nenghai Yu}.} \bibinfo{year}{2023}\natexlab{}.
\newblock \showarticletitle{Discop: Provably Secure Steganography in Practice Based on "Distribution Copies"}. In \bibinfo{booktitle}{\emph{2023 IEEE Symposium on Security and Privacy (SP)}}. \bibinfo{pages}{2238--2255}.
\newblock
\href{https://doi.org/10.1109/SP46215.2023.10179287}{doi:\nolinkurl{10.1109/SP46215.2023.10179287}}


\bibitem[enodari(2018)]%
        {zwsp-steg}
\bibfield{author}{\bibinfo{person}{enodari}.} \bibinfo{year}{2018}\natexlab{}.
\newblock \bibinfo{title}{zwsp-steg-py: Zero-Width Space Steganography}.
\newblock \bibinfo{howpublished}{GitHub repository}.
\newblock
\urldef\tempurl%
\url{https://github.com/enodari/zwsp-steg-py}
\showURL{%
\tempurl}


\bibitem[{Google Threat Intelligence Group}(2025)]%
        {gtig2025adversarial}
\bibfield{author}{\bibinfo{person}{{Google Threat Intelligence Group}}.} \bibinfo{year}{2025}\natexlab{}.
\newblock \bibinfo{title}{{Advances in Threat Actor Usage of AI Tools}}.
\newblock \bibinfo{howpublished}{Google Cloud Blog}.
\newblock
\urldef\tempurl%
\url{https://cloud.google.com/blog/topics/threat-intelligence/threat-actor-usage-of-ai-tools}
\showURL{%
\tempurl}


\bibitem[Greshake et~al\mbox{.}(2023)]%
        {greshake2023indirect}
\bibfield{author}{\bibinfo{person}{Kai Greshake}, \bibinfo{person}{Sahar Abdelnabi}, \bibinfo{person}{Shailesh Mishra}, \bibinfo{person}{Christoph Endres}, \bibinfo{person}{Thorsten Holz}, {and} \bibinfo{person}{Mario Fritz}.} \bibinfo{year}{2023}\natexlab{}.
\newblock \showarticletitle{Not What You've Signed Up For: Compromising Real-World LLM-Integrated Applications with Indirect Prompt Injection}. In \bibinfo{booktitle}{\emph{Proc. of the 16th ACM Workshop on Artificial Int. and Sec.}} (Copenhagen, Denmark) \emph{(\bibinfo{series}{AISec '23})}. \bibinfo{publisher}{Association for Computing Machinery}, \bibinfo{address}{New York, NY, USA}, \bibinfo{pages}{79–90}.
\newblock
\showISBNx{9798400702600}
\href{https://doi.org/10.1145/3605764.3623985}{doi:\nolinkurl{10.1145/3605764.3623985}}


\bibitem[Gupta et~al\mbox{.}(2023)]%
        {gupta2023threatgpt}
\bibfield{author}{\bibinfo{person}{Maanak Gupta}, \bibinfo{person}{Charankumar Akiri}, \bibinfo{person}{Kshitiz Aryal}, \bibinfo{person}{Eli Parker}, {and} \bibinfo{person}{Lopamudra Praharaj}.} \bibinfo{year}{2023}\natexlab{}.
\newblock \showarticletitle{From {ChatGPT} to {ThreatGPT}: Impact of Generative {AI} in Cybersecurity and Privacy}.
\newblock \bibinfo{journal}{\emph{IEEE Access}}  \bibinfo{volume}{11} (\bibinfo{year}{2023}), \bibinfo{pages}{80218--80245}.
\newblock
\href{https://doi.org/10.1109/ACCESS.2023.3300381}{doi:\nolinkurl{10.1109/ACCESS.2023.3300381}}


\bibitem[Gurman(2023)]%
        {gurman2023samsung}
\bibfield{author}{\bibinfo{person}{Mark Gurman}.} \bibinfo{year}{2023}\natexlab{}.
\newblock \bibinfo{title}{{Samsung Bans Generative AI Use by Staff After ChatGPT Data Leak}}.
\newblock \bibinfo{howpublished}{Bloomberg}.
\newblock
\urldef\tempurl%
\url{https://www.bloomberg.com/news/articles/2023-05-02/samsung-bans-chatgpt-and-other-generative-ai-use-by-staff-after-leak}
\showURL{%
\tempurl}


\bibitem[Han et~al\mbox{.}(2018)]%
        {han2018deception}
\bibfield{author}{\bibinfo{person}{Xiao Han}, \bibinfo{person}{Nizar Kheir}, {and} \bibinfo{person}{Davide Balzarotti}.} \bibinfo{year}{2018}\natexlab{}.
\newblock \showarticletitle{{Deception Techniques in Computer Security: A Research Perspective}}.
\newblock \bibinfo{journal}{\emph{Comput. Surveys}} \bibinfo{volume}{51}, \bibinfo{number}{4} (\bibinfo{year}{2018}), \bibinfo{pages}{1--36}.
\newblock
\href{https://doi.org/10.1145/3214305}{doi:\nolinkurl{10.1145/3214305}}


\bibitem[Hellmeier et~al\mbox{.}(2025)]%
        {hellmeier2025innamark}
\bibfield{author}{\bibinfo{person}{Malte Hellmeier}, \bibinfo{person}{Hendrik Norkowski}, \bibinfo{person}{Ernst-Christoph Schrewe}, \bibinfo{person}{Haydar Qarawlus}, {and} \bibinfo{person}{Falk Howar}.} \bibinfo{year}{2025}\natexlab{}.
\newblock \showarticletitle{{Innamark: A Whitespace Replacement Information-Hiding Method}}.
\newblock \bibinfo{journal}{\emph{{IEEE Access}}}  \bibinfo{volume}{13} (\bibinfo{year}{2025}), \bibinfo{pages}{123120--123135}.
\newblock
\href{https://doi.org/10.1109/ACCESS.2025.3583591}{doi:\nolinkurl{10.1109/ACCESS.2025.3583591}}


\bibitem[Homoliak et~al\mbox{.}(2019)]%
        {homoliak2019insiderthreat}
\bibfield{author}{\bibinfo{person}{Ivan Homoliak}, \bibinfo{person}{Flavio Toffalini}, \bibinfo{person}{Juan Guarnizo}, \bibinfo{person}{Yuval Elovici}, {and} \bibinfo{person}{Mart{\'i}n Ochoa}.} \bibinfo{year}{2019}\natexlab{}.
\newblock \showarticletitle{Insight Into Insiders and {IT}: A Survey of Insider Threat Taxonomies, Analysis, Modeling, and Countermeasures}.
\newblock \bibinfo{journal}{\emph{Comput. Surveys}} \bibinfo{volume}{52}, \bibinfo{number}{2}, Article \bibinfo{articleno}{30} (\bibinfo{year}{2019}), \bibinfo{numpages}{40}~pages.
\newblock
\href{https://doi.org/10.1145/3303771}{doi:\nolinkurl{10.1145/3303771}}


\bibitem[Kaptchuk et~al\mbox{.}(2021)]%
        {kaptchuk2021meteor}
\bibfield{author}{\bibinfo{person}{Gabriel Kaptchuk}, \bibinfo{person}{Tushar~M. Jois}, \bibinfo{person}{Matthew Green}, {and} \bibinfo{person}{Aviel~D. Rubin}.} \bibinfo{year}{2021}\natexlab{}.
\newblock \showarticletitle{Meteor: Cryptographically Secure Steganography for Realistic Distributions}. In \bibinfo{booktitle}{\emph{Proceedings of the 2021 ACM SIGSAC Conference on Computer and Communications Security}} (Virtual Event, Republic of Korea) \emph{(\bibinfo{series}{CCS '21})}. \bibinfo{publisher}{Association for Computing Machinery}, \bibinfo{address}{New York, NY, USA}, \bibinfo{pages}{1529–1548}.
\newblock
\showISBNx{9781450384544}
\href{https://doi.org/10.1145/3460120.3484550}{doi:\nolinkurl{10.1145/3460120.3484550}}


\bibitem[Kirchenbauer et~al\mbox{.}(2023)]%
        {kirchenbauer2023watermark}
\bibfield{author}{\bibinfo{person}{John Kirchenbauer}, \bibinfo{person}{Jonas Geiping}, \bibinfo{person}{Yuxin Wen}, \bibinfo{person}{Jonathan Katz}, \bibinfo{person}{Ian Miers}, {and} \bibinfo{person}{Tom Goldstein}.} \bibinfo{year}{2023}\natexlab{}.
\newblock \showarticletitle{A Watermark for Large Language Models}. In \bibinfo{booktitle}{\emph{Proceedings of the 40th International Conference on Machine Learning}} \emph{(\bibinfo{series}{Proceedings of Machine Learning Research}, Vol.~\bibinfo{volume}{202})}, \bibfield{editor}{\bibinfo{person}{Andreas Krause}, \bibinfo{person}{Emma Brunskill}, \bibinfo{person}{Kyunghyun Cho}, \bibinfo{person}{Barbara Engelhardt}, \bibinfo{person}{Sivan Sabato}, {and} \bibinfo{person}{Jonathan Scarlett}} (Eds.). \bibinfo{publisher}{PMLR}, \bibinfo{pages}{17061--17084}.
\newblock
\urldef\tempurl%
\url{https://proceedings.mlr.press/v202/kirchenbauer23a.html}
\showURL{%
\tempurl}


\bibitem[{Microsoft Threat Intelligence} and {OpenAI}(2024)]%
        {microsoft2024stayingahead}
\bibfield{author}{\bibinfo{person}{{Microsoft Threat Intelligence}} {and} \bibinfo{person}{{OpenAI}}.} \bibinfo{year}{2024}\natexlab{}.
\newblock \bibinfo{title}{{Staying Ahead of Threat Actors in the Age of AI}}.
\newblock \bibinfo{howpublished}{Microsoft Security Blog}.
\newblock
\urldef\tempurl%
\url{https://www.microsoft.com/en-us/security/blog/2024/02/14/staying-ahead-of-threat-actors-in-the-age-of-ai/}
\showURL{%
\tempurl}


\bibitem[Mishra et~al\mbox{.}(2025)]%
        {mishra2025pii}
\bibfield{author}{\bibinfo{person}{Kushagra Mishra}, \bibinfo{person}{Harsh Pagare}, {and} \bibinfo{person}{Kanhaiya Sharma}.} \bibinfo{year}{2025}\natexlab{}.
\newblock \showarticletitle{A hybrid rule-based {NLP} and machine learning approach for {PII} detection and anonymization in financial documents}.
\newblock \bibinfo{journal}{\emph{Scientific Reports}}  \bibinfo{volume}{15} (\bibinfo{year}{2025}), \bibinfo{pages}{22729}.
\newblock
\href{https://doi.org/10.1038/s41598-025-04971-9}{doi:\nolinkurl{10.1038/s41598-025-04971-9}}


\bibitem[Newman and Burgess(2025)]%
        {newman2025airansomware}
\bibfield{author}{\bibinfo{person}{Lily~Hay Newman} {and} \bibinfo{person}{Matt Burgess}.} \bibinfo{year}{2025}\natexlab{}.
\newblock \bibinfo{title}{The Era of {AI}-Generated Ransomware Has Arrived}.
\newblock \bibinfo{howpublished}{WIRED}.
\newblock
\urldef\tempurl%
\url{https://www.wired.com/story/the-era-of-ai-generated-ransomware-has-arrived/}
\showURL{%
\tempurl}


\bibitem[{OWASP Foundation}(2025)]%
        {owasp2025llmtop10}
\bibfield{author}{\bibinfo{person}{{OWASP Foundation}}.} \bibinfo{year}{2025}\natexlab{}.
\newblock \bibinfo{title}{{OWASP Top 10 for Large Language Model Applications 2025}}.
\newblock
\urldef\tempurl%
\url{https://genai.owasp.org/resource/owasp-top-10-for-llm-applications-2025/}
\showURL{%
\tempurl}


\bibitem[Putrevu et~al\mbox{.}(2024)]%
        {putrevu2024comprehensive}
\bibfield{author}{\bibinfo{person}{Mohan~Anand Putrevu}, \bibinfo{person}{Hrushikesh Chunduri}, \bibinfo{person}{Venkata Sai~Charan Putrevu}, {and} \bibinfo{person}{Sandeep~K Shukla}.} \bibinfo{year}{2024}\natexlab{}.
\newblock \showarticletitle{A comprehensive analysis of machine learning based file trap selection methods to detect crypto ransomware}.
\newblock \bibinfo{journal}{\emph{arXiv preprint arXiv:2409.11428}} (\bibinfo{year}{2024}).
\newblock


\bibitem[Radford et~al\mbox{.}(2019)]%
        {radford2019gpt2}
\bibfield{author}{\bibinfo{person}{Alec Radford}, \bibinfo{person}{Jeffrey Wu}, \bibinfo{person}{Rewon Child}, \bibinfo{person}{David Luan}, \bibinfo{person}{Dario Amodei}, {and} \bibinfo{person}{Ilya Sutskever}.} \bibinfo{year}{2019}\natexlab{}.
\newblock \bibinfo{booktitle}{\emph{Language Models are Unsupervised Multitask Learners}}.
\newblock \bibinfo{type}{{T}echnical {R}eport}. \bibinfo{institution}{OpenAI}.
\newblock
\urldef\tempurl%
\url{https://cdn.openai.com/better-language-models/language_models_are_unsupervised_multitask_learners.pdf}
\showURL{%
\tempurl}


\bibitem[Raz et~al\mbox{.}(2025)]%
        {raz2025ransomware30}
\bibfield{author}{\bibinfo{person}{Md Raz}, \bibinfo{person}{Meet Udeshi}, \bibinfo{person}{Venkata Sai~Charan Putrevu}, \bibinfo{person}{Prashanth Krishnamurthy}, \bibinfo{person}{Farshad Khorrami}, {and} \bibinfo{person}{Ramesh Karri}.} \bibinfo{year}{2025}\natexlab{}.
\newblock \bibinfo{title}{Ransomware 3.0: Self-Composing and {LLM}-Orchestrated}.
\newblock \bibinfo{howpublished}{arXiv preprint}.
\newblock
\showeprint[arxiv]{2508.20444}~[cs.CR]
\href{https://doi.org/10.48550/arXiv.2508.20444}{doi:\nolinkurl{10.48550/arXiv.2508.20444}}


\bibitem[Rizzo et~al\mbox{.}(2016)]%
        {rizzo2016}
\bibfield{author}{\bibinfo{person}{Stefano~Giovanni Rizzo}, \bibinfo{person}{Flavio Bertini}, {and} \bibinfo{person}{Danilo Montesi}.} \bibinfo{year}{2016}\natexlab{}.
\newblock \showarticletitle{Content-preserving Text Watermarking through Unicode Homoglyph Substitution}. In \bibinfo{booktitle}{\emph{Proceedings of the 20th International Database Engineering \& Applications Symposium}} (Montreal, QC, Canada) \emph{(\bibinfo{series}{IDEAS '16})}. \bibinfo{publisher}{Association for Computing Machinery}, \bibinfo{address}{New York, NY, USA}, \bibinfo{pages}{97–104}.
\newblock
\showISBNx{9781450341189}
\href{https://doi.org/10.1145/2938503.2938510}{doi:\nolinkurl{10.1145/2938503.2938510}}


\bibitem[Shen et~al\mbox{.}(2020)]%
        {shen2020saac}
\bibfield{author}{\bibinfo{person}{Jiaming Shen}, \bibinfo{person}{Heng Ji}, {and} \bibinfo{person}{Jiawei Han}.} \bibinfo{year}{2020}\natexlab{}.
\newblock \showarticletitle{Near-imperceptible Neural Linguistic Steganography via Self-Adjusting Arithmetic Coding}. In \bibinfo{booktitle}{\emph{Proceedings of the 2020 Conference on Empirical Methods in Natural Language Processing (EMNLP)}}, \bibfield{editor}{\bibinfo{person}{Bonnie Webber}, \bibinfo{person}{Trevor Cohn}, \bibinfo{person}{Yulan He}, {and} \bibinfo{person}{Yang Liu}} (Eds.). \bibinfo{publisher}{Association for Computational Linguistics}, \bibinfo{address}{Online}, \bibinfo{pages}{303--313}.
\newblock
\href{https://doi.org/10.18653/v1/2020.emnlp-main.22}{doi:\nolinkurl{10.18653/v1/2020.emnlp-main.22}}


\bibitem[Spitzner(2003)]%
        {spitzner2003honeypots}
\bibfield{author}{\bibinfo{person}{Lance Spitzner}.} \bibinfo{year}{2003}\natexlab{}.
\newblock \bibinfo{booktitle}{\emph{{Honeypots: Tracking Hackers}}}.
\newblock \bibinfo{publisher}{Addison-Wesley Professional}.
\newblock
\showISBNx{978-0-321-10895-1}


\bibitem[Swain(2026)]%
        {swain2026cisa}
\bibfield{author}{\bibinfo{person}{Gyana Swain}.} \bibinfo{year}{2026}\natexlab{}.
\newblock \bibinfo{title}{{CISA} Chief Uploaded Sensitive Government Files to Public {ChatGPT}}.
\newblock
\urldef\tempurl%
\url{https://www.csoonline.com/article/4124320/cisa-chief-uploaded-sensitive-government-files-to-public-chatgpt.html}
\showURL{%
\tempurl}


\bibitem[Taleby~Ahvanooey et~al\mbox{.}(2018)]%
        {ahvanooey2018aitsteg}
\bibfield{author}{\bibinfo{person}{Milad Taleby~Ahvanooey}, \bibinfo{person}{Qianmu Li}, \bibinfo{person}{Jun Hou}, \bibinfo{person}{Hassan Dana~Mazraeh}, {and} \bibinfo{person}{Jing Zhang}.} \bibinfo{year}{2018}\natexlab{}.
\newblock \showarticletitle{AITSteg: An Innovative Text Steganography Technique for Hidden Transmission of Text Message via Social Media}.
\newblock \bibinfo{journal}{\emph{IEEE Access}}  \bibinfo{volume}{6} (\bibinfo{year}{2018}), \bibinfo{pages}{65981--65995}.
\newblock
\href{https://doi.org/10.1109/ACCESS.2018.2866063}{doi:\nolinkurl{10.1109/ACCESS.2018.2866063}}


\bibitem[{Thinkst Applied Research}(2015)]%
        {thinkst2015canary}
\bibfield{author}{\bibinfo{person}{{Thinkst Applied Research}}.} \bibinfo{year}{2015}\natexlab{}.
\newblock \bibinfo{title}{{Thinkst Canary}}.
\newblock
\urldef\tempurl%
\url{https://canary.tools/}
\showURL{%
\tempurl}


\bibitem[Udeshi et~al\mbox{.}(2025)]%
        {udeshi2025samosa}
\bibfield{author}{\bibinfo{person}{Meet Udeshi}, \bibinfo{person}{Venkata Sai~Charan Putrevu}, \bibinfo{person}{Prashanth Krishnamurthy}, \bibinfo{person}{Ramesh Karri}, {and} \bibinfo{person}{Farshad Khorrami}.} \bibinfo{year}{2025}\natexlab{}.
\newblock \bibinfo{title}{{SaMOSA}: Sandbox for Malware Orchestration and Side-Channel Analysis}.
\newblock
\showeprint[arxiv]{2508.14261}~[cs.CR]
\href{https://doi.org/10.48550/arXiv.2508.14261}{doi:\nolinkurl{10.48550/arXiv.2508.14261}}


\bibitem[{Wikimedia Foundation}(2023)]%
        {wikimedia2023wikipedia}
\bibfield{author}{\bibinfo{person}{{Wikimedia Foundation}}.} \bibinfo{year}{2023}\natexlab{}.
\newblock \bibinfo{title}{Wikipedia Dataset}.
\newblock \bibinfo{howpublished}{Hugging Face Datasets}.
\newblock
\urldef\tempurl%
\url{https://huggingface.co/datasets/wikimedia/wikipedia}
\showURL{%
\tempurl}


\bibitem[Yuill et~al\mbox{.}(2004)]%
        {yuill2004honeyfiles}
\bibfield{author}{\bibinfo{person}{J. Yuill}, \bibinfo{person}{M. Zappe}, \bibinfo{person}{D. Denning}, {and} \bibinfo{person}{F. Feer}.} \bibinfo{year}{2004}\natexlab{}.
\newblock \showarticletitle{Honeyfiles: deceptive files for intrusion detection}. In \bibinfo{booktitle}{\emph{Proceedings from the Fifth Annual IEEE SMC Information Assurance Workshop, 2004.}} \bibinfo{pages}{116--122}.
\newblock
\href{https://doi.org/10.1109/IAW.2004.1437806}{doi:\nolinkurl{10.1109/IAW.2004.1437806}}


\bibitem[Zhang and Thing(2021)]%
        {zhang2021deception}
\bibfield{author}{\bibinfo{person}{Li Zhang} {and} \bibinfo{person}{Vrizlynn L.~L. Thing}.} \bibinfo{year}{2021}\natexlab{}.
\newblock \showarticletitle{{Three Decades of Deception Techniques in Active Cyber Defense---Retrospect and Outlook}}.
\newblock \bibinfo{journal}{\emph{Computers \& Security}}  \bibinfo{volume}{106} (\bibinfo{year}{2021}), \bibinfo{pages}{102288}.
\newblock
\href{https://doi.org/10.1016/j.cose.2021.102288}{doi:\nolinkurl{10.1016/j.cose.2021.102288}}


\bibitem[Zhang et~al\mbox{.}(2021)]%
        {zhang2021adg}
\bibfield{author}{\bibinfo{person}{Siyu Zhang}, \bibinfo{person}{Zhongliang Yang}, \bibinfo{person}{Jinshuai Yang}, {and} \bibinfo{person}{Yongfeng Huang}.} \bibinfo{year}{2021}\natexlab{}.
\newblock \showarticletitle{Provably Secure Generative Linguistic Steganography}. In \bibinfo{booktitle}{\emph{Findings of the Association for Computational Linguistics: ACL-IJCNLP 2021}}, \bibfield{editor}{\bibinfo{person}{Chengqing Zong}, \bibinfo{person}{Fei Xia}, \bibinfo{person}{Wenjie Li}, {and} \bibinfo{person}{Roberto Navigli}} (Eds.). \bibinfo{publisher}{Association for Computational Linguistics}, \bibinfo{address}{Online}, \bibinfo{pages}{3046--3055}.
\newblock
\href{https://doi.org/10.18653/v1/2021.findings-acl.268}{doi:\nolinkurl{10.18653/v1/2021.findings-acl.268}}


\bibitem[Ziegler et~al\mbox{.}(2019)]%
        {ziegler2019neural}
\bibfield{author}{\bibinfo{person}{Zachary Ziegler}, \bibinfo{person}{Yuntian Deng}, {and} \bibinfo{person}{Alexander Rush}.} \bibinfo{year}{2019}\natexlab{}.
\newblock \showarticletitle{Neural Linguistic Steganography}. In \bibinfo{booktitle}{\emph{Proceedings of the 2019 Conference on Empirical Methods in Natural Language Processing and the 9th International Joint Conference on Natural Language Processing (EMNLP-IJCNLP)}}, \bibfield{editor}{\bibinfo{person}{Kentaro Inui}, \bibinfo{person}{Jing Jiang}, \bibinfo{person}{Vincent Ng}, {and} \bibinfo{person}{Xiaojun Wan}} (Eds.). \bibinfo{publisher}{Association for Computational Linguistics}, \bibinfo{address}{Hong Kong, China}, \bibinfo{pages}{1210--1215}.
\newblock
\href{https://doi.org/10.18653/v1/D19-1115}{doi:\nolinkurl{10.18653/v1/D19-1115}}


\end{thebibliography}

\end{document}